\documentclass[showpacs,amsmath,nofootinbib,amssymb,twocolumn,superscriptaddress]{revtex4-1}
\usepackage{array}
\usepackage{color}
\usepackage{dcolumn}
\usepackage{bm}
\usepackage{epsf}
\usepackage{graphicx}
\usepackage{amsmath} 
\usepackage{dsfont} 
\usepackage{tikz}
\usepackage{hyperref}

\newcommand{\beq}{\begin{equation}}
\newcommand{\eeq}{\end{equation}}
\newcommand{\bea}{\begin{eqnarray}}
\newcommand{\eea}{\end{eqnarray}}

\newcommand\eqn[1]{(\ref{#1})}      
\newcommand\Eqn[1]{Eq.~(\ref{#1})}  

\newcommand\lp{\left(}
\newcommand\rp{\right)}

\newcommand\elq{\ell_q}

\begin{document}
\topmargin-0.5in
\textheight 8.7in 
\bibliographystyle{apsrev}
\title{Neutrino propagation  in binary neutron star mergers in presence of nonstandard interactions}
\author{Am\'elie Chatelain}
\email{chatelai@apc.in2p3.fr}
\affiliation{Astro-Particule et Cosmologie (APC), CNRS UMR 7164, Universit\'e Denis Diderot,\\ 10, rue Alice Domon et L\'eonie Duquet, 75205 Paris Cedex 13, France.}

\author{Maria Cristina Volpe}
\email{volpe@apc.in2p3.fr}
\affiliation{Astro-Particule et Cosmologie (APC), CNRS UMR 7164, Universit\'e Denis Diderot,\\ 10, rue Alice Domon et L\'eonie Duquet, 75205 Paris Cedex 13, France.}

\begin{abstract}
We explore the impact  of nonstandard interactions on neutrino propagation in accretion disks around binary neutron star mergers remnants. 
We show flavor evolution can be significantly modified even for values of the nonstandard couplings well below current bounds.
We demonstrate the occurrence of I resonances as synchronized MSW phenomena and show that intricate conversion patterns might appear depending on the nonstandard interaction parameters. We discuss the possible implications for nucleosynthesis.

\end{abstract}
\date{\today}


\maketitle
\section{Introduction}
The origin of heavy elements remains one of the key open questions in nuclear astrophysics.
Nucleosynthetic abundances  produced in the rapid neutron capture
process ($r$-process) are formed in dense neutron-rich environments \cite{Burbidge:1957vc} including
core-collapse supernovae,  accretion disks around black holes or binary compact systems. It was first recognized in \cite{Lattimer:1977cff} that $r$-process nuclei could be formed in neutron star matter.
The occurrence of a weak or of a strong $r$-process 
depends mainly upon the astrophysical conditions and the
properties of exotic nuclei. In particular, conditions for a 
strong r-process are met in neutron star mergers, whereas  elements with $A > 130 $ are not produced in core-collapse supernovae considered for long a favorite $r$-process site (see e.g. \cite{Qian:2013fsa,Arcones:2012wj}).

The recent observation of gravitational waves from a binary neutron star (BNS) merger event in coincident with a short gamma-ray-bursts and a kilonova constitute the first experimental evidence for r-process nucleosynthesis in such sites \cite{TheLIGOScientific:2017qsa, Pian:2017gtc}. Weak interactions and neutrinos bring the ejecta to being hot. The role of neutrino flavor evolution in these environments still needs to be fully assessed.  

Calculations of nucleosynthetic abundances of heavy elements show that dynamical ejecta can produce a strong r-process while a weak r-process can take place in neutrino driven winds \cite{Martin:2015hxa}. 
In fact, the presence of a significant amount of neutrinos  in neutrino-driven winds
 influences the build-up of heavy elements through the electron neutrino and antineutrino interactions 
with neutrons and protons respectively. 
Such interactions tend to be detrimental to the $r$-process since they reduce the number of available
neutrons. The occurrence of flavor conversion phenomena can produce swappings of the neutrino spectra and
modify the interaction rates that determine the electron fraction, as shown in numerous studies (see e.g. Refs.\cite{Balantekin:2004ug,Malkus:2012ts,Malkus:2015mda}).

Two decades of experiments have contributed to determine the neutrino squared-mass differences and the mixing angles responsible for vacuum oscillations and for some of the flavor conversion mechanisms in matter. 
Open questions remain concerning the absolute neutrino mass ordering, the neutrino nature (Dirac versus Majorana), the existence of sterile neutrinos and of CP violation in the leptonic sector.
The combined analysis
of available data indicate that the neutrino mass ordering is normal and the Dirac CP violating phase approximately $1.5 \pi$,  
although statistical significance
is still low \cite{Capozzi:2017ipn}. 

Flavor evolution in astrophysical environments  reveals
a variety of mechanisms due to the nonlinear many-body nature of neutrino propagation in presence of neutrino self-interactions.  
Regions of flavor instabilities close to the neutrinosphere are found  including 
collective phenomena such as the bipolar instability in core-collapse supernovae \cite{Chakraborty:2015tfa, Duan:2010bg},
or the matter-neutrino resonance  (MNR)  in neutron star mergers \cite{Malkus:2012ts}.

The presence of nonstandard  interactions can alter flavor conversion. Limits on nonstandard neutrino self-interactions are rather loose \cite{Bilenky:1999dn}, whereas scattering and oscillation experiments furnish tight bounds on nonstandard neutrino-matter interactions (NSI)  \cite{Davidson:2003ha,Biggio:2009nt,Ohlsson:2012kf}.
The first measurement of neutrino-nucleus coherent scattering provides interesting NSI constraints Ref.\cite{Akimov:2017ade}.
The existence of  NSI would modify the interpretation of oscillation experiments in particular for the inferred values of the squared-masses and the mixings 
and could furnish an explanation of observed anomalies.

Within a supernova core, flavor changing neutral current interactions would impact the scattering rates and the electron fraction, altering the infall
\cite{Amanik:2006ad}. Nonstandard four fermion neutrino self-interactions might produce flavor equilibration both in normal and inverted mass ordering \cite{Blennow:2008er} or could modify the neutronisation burst signal of a supernova explosion \cite{Das:2017iuj}.
Novel interactions can also produce resonant conversion nearby the neutrinosphere  and influence the $r$-process  in supernovae \cite{Nunokawa:1996tg}.
In particular, the  I resonance -- a Mikheev Smirnov Wolfenstein (MSW)-like resonance \cite{Wolfenstein:1977ue,Mikheev:1986gs} -- can take place  due to the cancellation between 
the matter and the NSI contributions to the neutrino Hamiltonian \cite{EstebanPretel:2007yu}. Refs. \cite{EstebanPretel:2009is,Stapleford:2016jgz}
have pointed out that the I location appears to be little affected by neutrino self-interactions.
Moreover Ref.\cite{Stapleford:2016jgz} has shown that NSI contributions can provide the necessary cancellation for the occurrence of MNR in supernovae.

Neutrino propagation in accretion disks of compact objects presents differences  with supernovae. 
Spherical symmetry breaking makes simulations demanding, even when the geometry of the 
neutrino emission at the neutrinospheres is the simplest. 
Detailed simulations of neutrino emission in accretion disks show that the
electron antineutrino luminosities are larger than the electron neutrino ones  and that the muon and tau neutrino luminosities are small.
Such excess of electron antineutrinos over neutrinos can produce the MNR when the matter and the neutrino self-interaction
potentials cancel  \cite{Malkus:2015mda}. 
A MNR is a series of MSW-like resonances \cite{Vaananen:2015hfa,Wu:2015fga,Malkus:2015mda} that can impact $r$-process nucleosynthesis 
 \cite{Malkus:2015mda}. 
A perturbative analysis unravels the conditions for nonlinear feedback that produces
the matching of the self-interaction and the matter contributions and maintains the MNR over several tens of kilometers.
On the other hand, such a matching is impeded by the geometrical factors radial dependence in the case of 
helicity coherence \cite{Chatelain:2016xva}. 
Conditions for the MNRs  can be met in detailed simulations of the disk around BNS merger remnants \cite{Zhu:2016mwa,Frensel:2016fge}. 
An analysis with non-stationary evolution shows that conversion can occur on very short timescales \cite{Wu:2017qpc}. These modes termed as {\it fast} have been first pointed out 
in the supernova context \cite{Sawyer:2015dsa}.

Our goal is to explore the role of nonstandard neutrino-matter interactions on the neutrino evolution in accretion disks around binary neutron star merger remnants.
First  we focus on the NSI impact on flavor evolution and discuss the I resonance. We shed a new light  on its mechanism and 
show that the neutrino self-interactions can produce I resonances as synchronized MSW effects. 
Moreover we present how NSI can modify both location and adiabaticity of the MNRs.
Our calculations are based on the matter density profiles and electron fraction taken from detailed astrophysical 
simulations of BNS remnants \cite{Perego:2014fma}. We discuss the effects of nonstandard interactions on the electron fraction $Y_e$, a key parameter for $r$-process nucleosynthesis in neutrino-driven winds, in the light of the study of Ref.\cite{Martin:2015hxa}.

The manuscript is structured as follows. Section II presents the model with NSI.
Numerical results on the flavor evolution for different sets of NSI parameters are given in Section III. The NSI effects on the I and MNR resonances are discussed.
Section IV is a conclusion.

\section{The model}
\subsection{Neutrino evolution equations in presence of nonstandard interactions}
\noindent
The evolution of a system of neutrinos and antineutrinos in an astrophysical environment is governed by the Liouville Von-Neumann equations  (we will use $\hbar=c=1$) 
\begin{align}\label{e:LVN}
i \dot{\rho}  = [h, \rho], \, \, \, \, \, \, \, \, \, \, \, \, \, \, 
 i \dot{\bar{\rho}}  = [\bar{h}, \bar{\rho}] ,
 \end{align}
where ${\rho} $ and $\bar{\rho}$ are single-particle density matrices, $h$ and $\bar{h}$ mean-field Hamiltonians for neutrinos and antineutrinos respectively.
For a detailed derivation of Eqs.(\ref{e:LVN}) see e.g.  \cite{Serreau:2014cfa}.
The mean-field equations \eqref{e:LVN} correspond to the lowest order truncation of the Born-Bogoliubov-Green-Kirkwood-Yvon hierarchy
for the two-point correlators  \cite{Volpe:2013jgr} 
\begin{align}\label{e:rho}
 \rho_{ij}(t,\vec q,-,\vec q\,'\!,-)&=\langle a^\dagger_j(t,\vec q,-)a_i(t,\vec q,-)\rangle,\\ \nonumber
 \bar\rho_{ij}(t,\vec q,+,\vec q\,'\!,+)&=\langle a^\dagger_i(t,\vec q,+)a_j(t,\vec q,+)\rangle,
\end{align}
that depend on time, on particle momentum $\vec{q}$ and on positive ($h=+$) or negative ($h=-$) helicity states.
The labelling $i,\,j$ are either  mass or flavor indices.
The creation and annihilation operators $a^{\dagger}$ ($b^{\dagger}$)  and 
$a$ ($b$) for neutrinos (antineutrinos) satisfy the nonzero equal-time anti-commutation relations.
The diagonal elements ($i = j$) are the expectation values of the number operator, while the non-diagonal ones ($i \neq j $) include the decoherence due to the neutrino mixing. In the following the explicit dependence on the momentum and helicity variables, as well as the time dependence will not be shown systematically to simplify notations.

Since neutrinos propagate through an astrophysical background,
the mean-field Hamiltonians include the neutrino charged- and neutral-current interactions with the particles composing the medium, usually electrons, protons and neutrons, as we will be considering in the present work. Therefore $h$ is given by
\begin{align}
\label{e:h}
h&=h_0 +h_{\rm mat}+ h_{\nu\nu},
\end{align}
where the first term corresponds to the vacuum Hamiltonian, the second to the neutrino standard and nonstandard interactions with matter and the last one to neutrino self-interactions.
The same expression holds for $\bar{h}$ with a minus sign for the $h_0$ contribution.  
In the flavor basis, the vacuum term reads 
 \beq\label{e:hvac}
h_0 = \mathrm{U} h_{\rm vac} \mathrm{U}^{\dagger}, 
\eeq
with $h_{\rm vac} = diag(E_i)$, $E_{i = 1, N_f}$ being the eigenenergies of the propagation eigenstates with $N_f$ the number of neutrino flavors.
The quantity $\mathrm{U}$ is the Pontecorvo-Maki-Nakagawa-Sakata (PMNS) $N_f \times N_f$ unitary matrix relating the mass to the flavor basis \cite{Maki:1962mu}.

As for the matter term, 
it comprises the standard contribution from neutrino-electron charged currents\footnote{We remind that the standard neutrino-matter neutral current contributions are not included since they are proportional to the identity matrix and therefore do not produce flavor modifications.} and a nonstandard term related to neutrino-matter interactions
\begin{align}\label{e:hmat}
h_{\rm mat} &= h_{\rm CC} + h_{\rm NSI},
 \end{align}
 where $h_{\rm CC} = diag(V_{CC}, 0)$ and $V_{CC} = \sqrt{2} G_F \rho_e$, with $G_F$ the Fermi coupling constant and  $\rho_e $ the net electron number density.  
 Note that here anisotropic contributions to the matter Hamiltonian are not included\footnote{Such contributions are e.g. implemented in 
Ref. \cite{Chatelain:2016xva}. Also trace terms can be subtracted from the Hamiltonian whereas this is not possible  in presence of helicity coherence \cite{Chatelain:2016xva}. }.
 The nonstandard interaction Hamiltonian is 
 \begin{align}\label{e:hnsi}
h_{\rm NSI} = \sqrt{2} G_F \sum_f n_f \epsilon^f,
 \end{align}
where  a sum over the electron, down and up quark\footnote{The heavy quark content of the nucleon is neglected.} number densities is performed
($f = e, d, u $). The $\epsilon$ matrices correspond to the nonstandard interactions couplings, constrained by several observations  \cite{Davidson:2003ha,Biggio:2009nt,Ohlsson:2012kf,Akimov:2017ade}.
In the case of three neutrino flavors, these are  \cite{Biggio:2009nt}
\beq\label{e:NSIpar}
\left(\begin{tabular}{lll}
\, $ \mid\epsilon_{ee}\mid \, < \, 2.5 $ \, &   \, $\mid \epsilon_{e\mu} \mid \, < \, 0.21$ \,  & \, $\mid  \epsilon_{e\tau }\mid \,  < \, 1.7$\,  \\
 &\,  $\mid  \epsilon_{ \mu \mu}\mid \, < \, 0.046  $ \, & \, $\mid  \epsilon_{\mu \tau}\mid  \, <\,  0.21 $ \, \\
 & & \, $\mid  \epsilon_{\tau \tau } \mid \,  < \, 9.0$ \, 
\end{tabular} \right),
\eeq
if matter is composed only of protons and electrons (solar--like).  
One can see that the bounds on the NSI parameters are rather loose, with the exception of $ \epsilon_{ \mu \mu}$.

The third contribution in Eq.(\ref{e:h}) corresponds to the neutrino self-interaction Hamiltonian 
 \begin{align}\label{e:hnunu}
h_{\nu \nu} & = \sqrt{2}G_F \sum_{\alpha}  \int  (1 - \hat{q} \cdot \hat{p})  \\ \nonumber
&   \times \left[ \mathrm dn_{\nu_{\underline{\alpha}}} \rho_{\nu_{\underline{\alpha}}} ({\vec p}) - \mathrm dn_{\bar{\nu}_{\underline{\alpha}}}  \bar{\rho}_{ \bar{\nu}_{\underline{\alpha}}}({\vec p}) \right]  ,
 \end{align}
where the quantity $\mathrm dn_{\nu_{\underline{\alpha}}}$($\mathrm dn_{\bar{\nu}_{\underline{\alpha}}}$) denotes the differential number density of neutrinos (antineutrinos), the underline refers to the neutrinos initially born with $\alpha $ flavor at the neutrinosphere. 

\subsection{Two neutrino flavor evolution in binary neutron star mergers}
\noindent
We employ the theoretical framework of two-neutrino flavors and stationary evolution\footnote{From now on,  only the radial dependence of all quantities is retained
and not explicitly shown to simplify notations.}.
In the flavor basis the neutrino density matrix  Eq.\eqref{e:rho}  reads  
 \beq\label{e:rho2f}
\rho=  
\left(\begin{tabular}{cc}
$\rho_{ee}$  &   $\rho_{ex}$  \\
$\rho_{ex}^*$ & $\rho_{xx}$ 
\end{tabular} \right),
\eeq
and similarly for $\bar{\rho}$. 
The vacuum Hamiltonian Eq.\eqref{e:hvac} involves the PMNS matrix that for three flavors depends on three measured mixing angles
and three unknown CP violating phases (one Dirac- and two Majorana-type) \cite{Maki:1962mu}. 
In two flavors, these fundamental parameters reduce to one mixing angle $\theta$ (one phase as well in the case of Majorana neutrinos). Therefore  the vacuum contribution becomes
\beq\label{e:hvac2f}
h_0 = \omega
\left(\begin{tabular}{cc}
$- c_{2\theta} $ & $ s_{2\theta} $\\
$s_{2\theta}$ & $ c_{2\theta} $ \\
\end{tabular} \right),
\eeq
with $\omega = {\Delta m^2 \over{4E}} $, $\Delta m^2 = m^2_2 - m^2_1$ with $m_1, m_2$ the mass values of the mass eigenstates,  $E = q$ the neutrino energy, $s_{2\theta} = \sin 2 \theta$ and $c_{2\theta} = \cos 2 \theta$.

\begin{figure}
\includegraphics[scale=1]{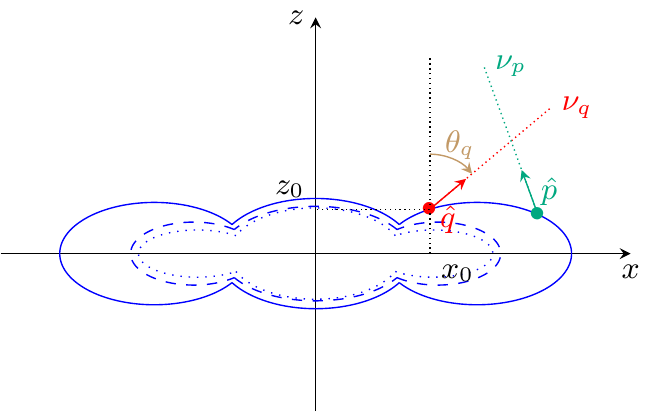}
\caption{Schematic view of our model. Neutrinos start free streaming at the neutrinospheres, shown as a blue solid (respectively dashed and dotted) line for $\nu_e$ (respectively $\bar{\nu}_e$ and $\nu_x$). The trajectory of a test neutrino $\nu_q$ is labelled by the coordinates of its emission point $\lp x_0, z_0 \rp$, and the angle $\theta_q$ between the direction of its momentum $\hat{q}$ and the $z$-axis. The test neutrino propagates in a background of matter and (anti)neutrinos $\nu_p$ of momentum $\hat{p}$.}\label{fig1} 
\end{figure}

For the standard matter Hamiltonian in Eq.(\ref{e:hmat}) we write
\begin{align}\label{e:hmat2f}
V_{CC} &= \lambda Y_e, 
 \end{align}
where $\lambda = \sqrt{2} G_F n_B$, with $n_B$ the baryon number density,  and $Y_e = \rho_e/(n+p)$ the electron fraction, with $n$ and $p$ the neutron and proton number densities, respectively.
As in Refs.\cite{Zhu:2016mwa,Frensel:2016fge,Chatelain:2016xva} our investigation is anchored to the detailed simulations in which the BNS merger remnant is a central object, lasts up to 200 ms and  has about 30 km radius.
We take information on the baryon number densities and electron fraction 
from cylindrical averages of detailed three-dimensional Newtonian simulations \cite{Perego:2014fma}.
In our  two-dimensional model neutrino propagate with an azimuthal symmetry axis from point  $(x_0, z_0)$, at the neutrinosphere following a straight line trajectory\footnote{In this work, we neglect the bending of the trajectory
due to strong gravitational fields.} characterized by a radial $r$ and an angular $\theta_q$ variables (Figure \ref{fig1}). 
Note that we approximate the neutrinospheres
as infinitely thin disks of radii $R_{\nu}$ that are flavor dependent, as done in Refs.\cite{Malkus:2012ts,Malkus:2015mda,Zhu:2016mwa,Frensel:2016fge,Chatelain:2016xva}.

In two-flavors, by retaining onlu the nonstandard contribution Eq.(\ref{e:hnsi}-\ref{e:NSIpar}) with loosest constraints, we get for the $\epsilon$ matrix
 \begin{align}\label{e:nsi2f}
\left(\begin{tabular}{ll}
 \, $ \mid\epsilon_{ee}\mid \, <  \, 2.5 $  \,  &  \,  $\mid  \epsilon_{e\tau}  \mid  \,  <  \, 1.7$  \, \\
  &  \, $\mid  \epsilon_{\tau \tau } \mid  \,  < \,  9.0 \, $ 
\end{tabular} \right).
\end{align}
We rewrite the NSI potential Eq.(\ref{e:hnsi}) in terms of the fermion fraction $Y_f$. In fact, using the charge neutrality of the medium, we get the relation
\beq\label{e:Yf}
Y_f \equiv {n_f \over n_B},
\eeq
which for up and down quarks can be rewritten as $Y_d = 2 - Y_e$ and $Y_u = 1 + Y_e$. The NSI contribution is then
\beq\label{e:Vnsi}
h_{\rm NSI} = \sqrt{2} G_F n_B \left[ Y_e \epsilon^e +  (1 + Y_e) \epsilon^u + (2 - Y_e) \epsilon^d  \right].
\eeq
Finally we follow Ref. \cite{Stapleford:2016jgz} and impose the requirement that, at the MSW resonance in the Sun, with an electron fraction $Y_{\odot} \approx 0.7$, the NSI contribution should vanish as no effect has been observed (see also \cite{Maltoni:2015kca}),
namely 
\beq\label{e:Yfsun}
 Y_{\odot} \delta \epsilon^e +  (1 + Y_{\odot} ) \delta \epsilon^u + (2 - Y_{\odot} ) \delta \epsilon^d   = 0,
\eeq
with $ \delta \epsilon^f =  \epsilon_{ee}^f -  \epsilon_{xx}^f $. This equation gives a relation between $ \delta \epsilon^e$ as a function of
$ \delta \epsilon^u,  \delta \epsilon^d$. The off-diagonal couplings $\epsilon^e_{ex}, \epsilon^u_{ex}, \epsilon^d_{ex}$ are fixed at the same value $\epsilon_0$. As a result the NSI Hamiltonian only depends on two NSI parameters,  the diagonal one $  \delta \epsilon^n$ and the off-diagonal 
$\epsilon_0$
\beq\label{e:nsi2par}
h_{\rm NSI} = \lambda 
\left(
 \begin{tabular}{ll}
$({{Y_{\odot} - Y_e} \over{Y_{\odot}}}) \delta \epsilon^n $ & $(3 + Y_e) \epsilon_0$ \\
$(3 + Y_e) \epsilon_0^*$  & 0 
\end{tabular} \right),
\eeq
with the constraints $\left| \delta \epsilon^n  \right|\lesssim \mathcal{O} \lp 10 \rp$ and $\left| \epsilon_0 \right| \lesssim \mathcal{O} \lp 1 \rp$.
For the neutrino self-interaction Hamiltonian Eq.(\ref{e:hnunu}) we assume, as done in previous works  \cite{Malkus:2012ts,Malkus:2015mda,Zhu:2016mwa,Frensel:2016fge,Chatelain:2016xva}, that
\begin{align}\label{e:st}
\rho_{\nu} (r, \vec{p}) = \rho_{\nu} (r, p) ,
\end{align}
namely that the angular dependence of the neutrino density matrix  is not retained. As a consequence, the neutrinos that are coupled by the self-interaction term have the same flavor history as the test neutrino. Assuming spherical and azimuthal symmetry of the neutrino emission at the neutrinosphere, this {\it ansatz} reduces to the single-angle approximation of the bulb-model \cite{Duan:2010bg}. 
Clearly, simulations implementing the full angular dependence of the density matrix Eq.\eqref{e:hnunu} are needed in the future to determine for example the role of decoherence in the flavor evolution. The linearized analysis of Ref.\cite{Wu:2017qpc} has included the angular dependence. We assume in our calculations that neutrinos are emitted as Fermi-Dirac distributions $f_{\nu_\alpha}$ with luminosities $L_{\nu_\alpha}$, average energies $\left\langle E_{\nu_\alpha} \right\rangle$ at the neutrinosphere with neutrinosphere radii $R_{\nu_\alpha}$ given in 
Table I. Table VII and Figures $14-16$ of Ref. \cite{Frensel:2016fge} show the current spread on $L_{\nu}$
and $\langle E_{\nu} \rangle$ according to available simulations of neutrino emission in binary neutron star mergers. Concerning the neutrino luminosities and average energies, these are stable for long times (see Ref. \cite{Perego:2014fma}). 
By using Eqs.(\ref{e:hnunu}) and (\ref{e:st}), the neutrino self-interaction term becomes\footnote{Note that here we show the full dependence on the variables for clarity.}   
\begin{widetext}
\begin{align}\label{e:hexpsa}
h_{\nu \nu} (r,q, \elq)  & =\sqrt{2} G_F  \sum_{\alpha=e,x}
   \int_{0}^{\infty}\! \mathrm dp   \left[ G_{\nu_{\alpha}} (r,\elq)\rho_{\nu_{\underline{\alpha}}}(r,p)  {{L_{\nu_{\underline{\alpha}}} f_{\nu_{\underline{\alpha}}}(p) }  \over{\pi^2 R^2_{\nu_\alpha}\left< E_{\nu_\alpha} \right>}} 
  -   \bar{\rho}_{ \bar{\nu}_{\underline{\alpha}}}(r,p) 
  G_{\bar{\nu}_{\alpha}} (r,\elq) {{L_{ \bar{\nu}_{\underline{\alpha}}} f_{\bar{\nu}_{\underline{\alpha}}}(p)}  \over{ \pi^2 R^2_{\bar{\nu}_\alpha}\left< E_{\bar{\nu}_\alpha} \right>}}  \right],
\end{align}
\end{widetext}
where the geometrical factor $G_{\nu_{\alpha}} $ reads
\begin{align}\label{e:geom}
G_{\nu_\alpha} (r,\ell_q)  = \int_{\Omega_{ \nu_{\alpha}}} \mathrm d \Omega(1 - \hat{q} \cdot \hat{p}) ,
\end{align}
with $\Omega_{ \nu_{\alpha}}$ the angular variables and similarly for $G_{\bar{\nu}_{\alpha}}$ for the antineutrinos, and $\ell_q = \lp \theta_q, x_0, z_0 \rp$.
The detailed derivation of the geometrical factors is given in Appendix B of Ref. \cite{Chatelain:2016xva} (see also Figures $2$ and $3$ of the same Reference). We introduce the unoscillated neutrino potential as 
\begin{align}\label{e:muunosc}
\mu & (r, \elq)  \equiv h^\text{unosc}_{\nu \nu,ee} (r, \elq) - h^\text{unosc}_{\nu \nu,xx} (r, \elq)\\ \nonumber
& \equiv \frac{\sqrt{2} G_F}{ \pi^2} \left[ G_{\nu_e} \lp r, \elq \rp \frac{L_{\nu_e}}{R^2_{\nu_e} \left\langle E_{\nu_e} \right\rangle} - G_{\bar{\nu}_e} \lp r, \elq \rp \frac{L_{\bar{\nu}_e}}{R^2_{\bar{\nu}_e} \left\langle E_{\bar{\nu}_e} \right\rangle} \right].
\end{align}

\begin{table} 
\center
\begin{tabular}{ccccccc}
~~~~~~~~~&  $\langle E_{\nu} \rangle$  ({\rm MeV}) & $L_{\nu}$   $(10^{51}~{\rm erg/s})$  & $R_{\nu}$  ({\rm km}) \hfill \\ \hline
$\nu_{e}$ &  $10.6$ & $15$  & 84 \\
$\bar{\nu}_{e}$ & $15.3$ & $30$ &  60 \\
$\nu_{x}$ & $17.3$ & $8$ &   58 \\
 \hline
\end{tabular}
\caption{Electron and nonelectron neutrino flavors : Average neutrino energies from Ref.\cite{Frensel:2016fge}, luminosities  from Ref.\cite{Perego:2014fma}. 
The last column furnishes the  outermost radii ({\rm km}) Ref. \cite{Frensel:2016fge}. Such values correspond to the neutrinospheres of a neutron star merger at 100 ms after the merging process. Please keep in mind that these luminosities have to be divided by two in Eq.(\ref{e:hexpsa}) because we consider there only neutrino emission in the half plane above the emission disk.}
\label{tab:fluxes}
\end{table}

\section{Impact of nonstandard interactions on neutrino flavor evolution}
\noindent
In order to investigate the role of NSI on the flavor evolution we have performed simulations 
by varying $\epsilon_0$ and $\delta \epsilon^n$ within the range given by relations (\ref{e:nsi2f}). 
We have explored a large set of trajectories with different emission points $(x_0, z_0)$ and  angles $\theta_q$ 
(Figure \ref{fig1})\footnote{Note that in our simulations $\phi_q$ is set to zero.}.
By analyzing the neutrino flavor evolution behaviors along numerous trajectories we have identified different regimes depending
on the NSI parameters. Here we take  some trajectories as typical examples  to illustrate the flavor mechanisms and their interplay we have
observed over the full set. As for the oscillation parameters we fix $\delta m^2 = 2.43 \times 10^{-3}~$eV$^2$ and $\sin^2 2 \theta = 0.087$ \cite{Patrignani:2016xqp} for the normal mass ordering, and $\delta m^2 = -2.38 \times 10^{-3}~$eV$^2$ and $\sin^2 2 \theta = 0.092$ for the inverted mass ordering.
We discuss the dependence of the results both on the normal and on the inverted mass ordering since the neutrino mass ordering has not been determined yet.

In this section, we have show examples with NSI parameters $\delta \epsilon^n \in \left[ -0.9, -0.7 \right]$. These are the parameters for which we observed the presence of the I resonance in most of the trajectories explored that were relevant for nucleosynthesis \cite{Martin:2015hxa}. Negative values of $\delta \epsilon^n$ with a greater absolute value lead to the disappearance of the I resonances, as the matter potential $V_M$ Eq.(\ref{e:Ires}) would always be negative on the region of space studied, and would also make the MNR further away. Negative values of $\delta \epsilon^n$ with a smaller absolute value would still present I resonances, but in a different region of space, and would also shift the MNRs. It is worth noting that positive values of $\delta \epsilon^n$ have also been considered as they can shift the MNR closer to the neutrinosphere. 

As for the value of $\epsilon_0$, we have restricted ourselves to values smaller than $10^{-3}$. Indeed, values larger than that creates oscillation patterns analogous to vacuum oscillations, but driven by the large matter off-diagonal element. These oscillations have a very short wavelength (shorter than a kilometer), and can start as soon as the neutrino propagation begins. Given that, in our calculations, we assume that neutrinos are free streaming, our results are reliable only if flavor conversions happen well outside the neutrinospheres, and therefore using larger values of $\epsilon_0$ would give unphysical results. These oscillations appearing because of a larger $\epsilon_0$ also have a large amplitude, making the behaviors difficult to analysis. For all these reasons, we chose to work with a value of $\epsilon_0$ well below the current experimental constraints. 

\subsection{New conditions for the I resonance}
The presence of NSI produces a new MSW-like resonance, called the inner (I) resonance \cite{EstebanPretel:2007yu}. 
Refs.\cite{EstebanPretel:2007yu,EstebanPretel:2009is,Stapleford:2016jgz} have shown that its occurrence is due to the matter terms only. In the present work 
we will be discussing two situations in which the I resonance occurs : {\it i)}  the self-interaction is sub-dominant, in accord with \cite{EstebanPretel:2007yu,EstebanPretel:2009is,Stapleford:2016jgz}; {\it ii)} the neutrino self-interaction dominates and leads to a
I resonance as a synchronized MSW mechanism. We explore this scenario using the $\mathrm{SU} \lp 2 \rp$ spin formalism.

\subsubsection{I resonance with negligible self-interaction}

\begin{figure*}
\includegraphics[width=.3\textwidth]{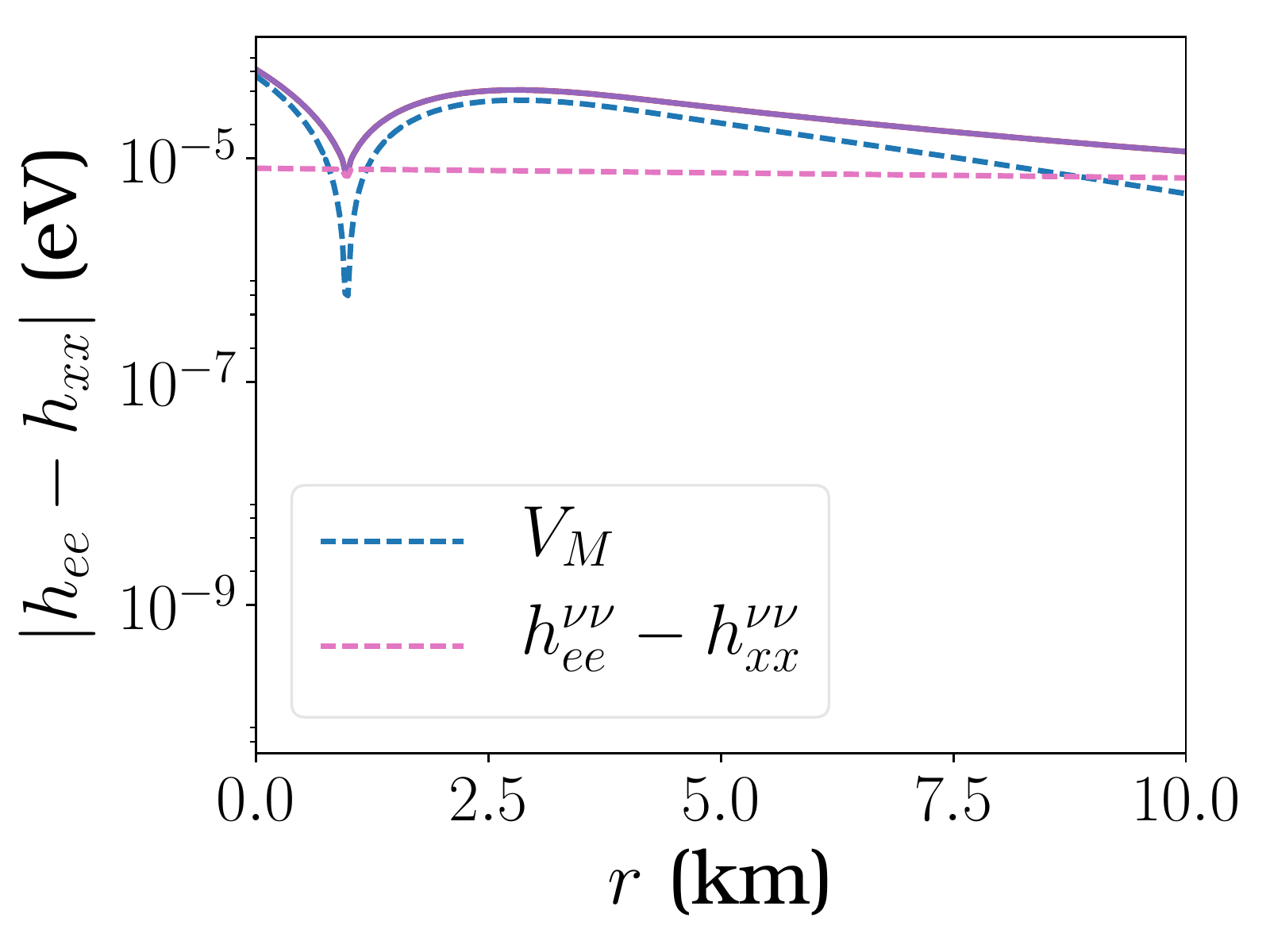}
\includegraphics[width=.3\textwidth]{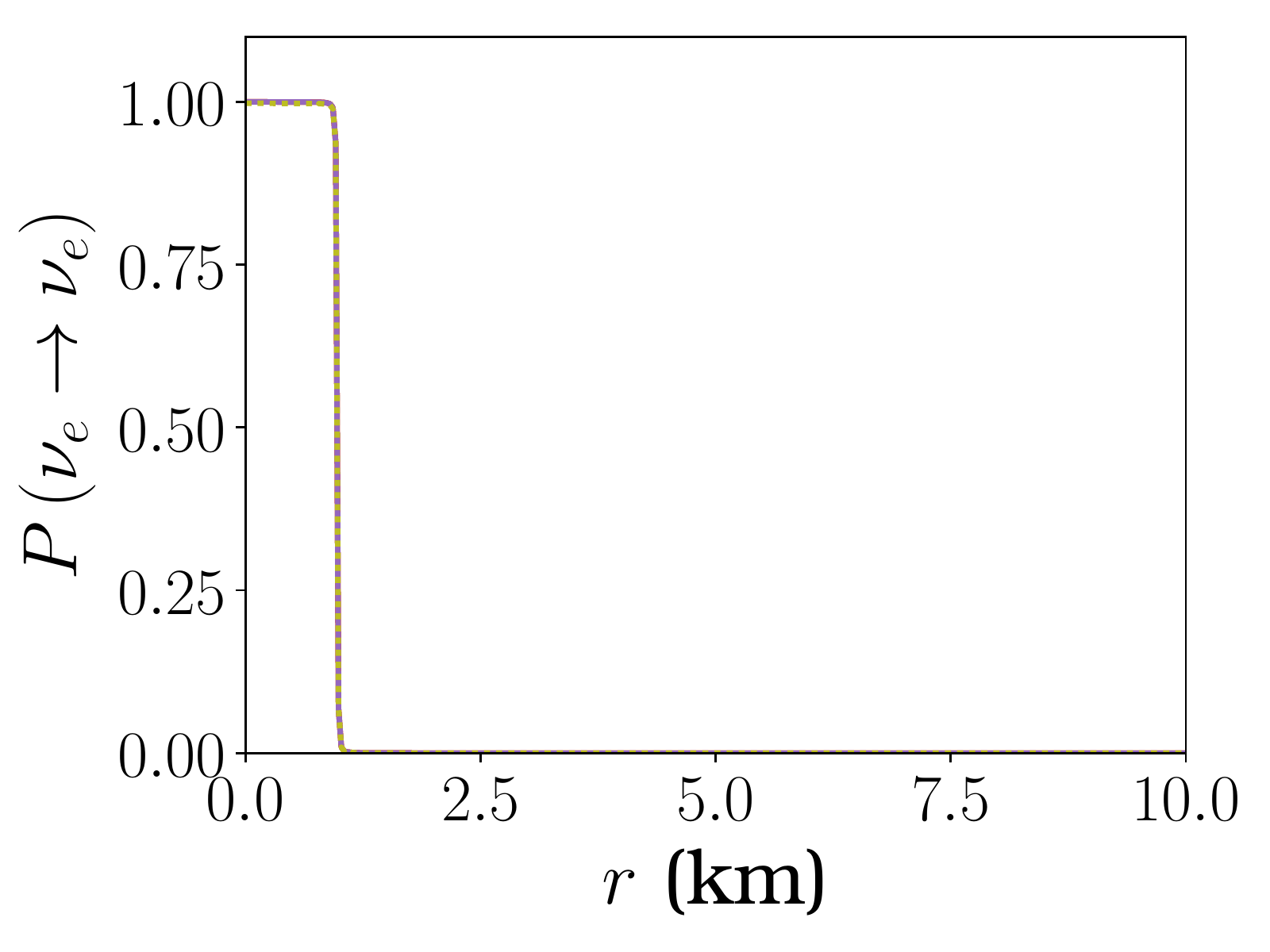}
\includegraphics[width=.3\textwidth]{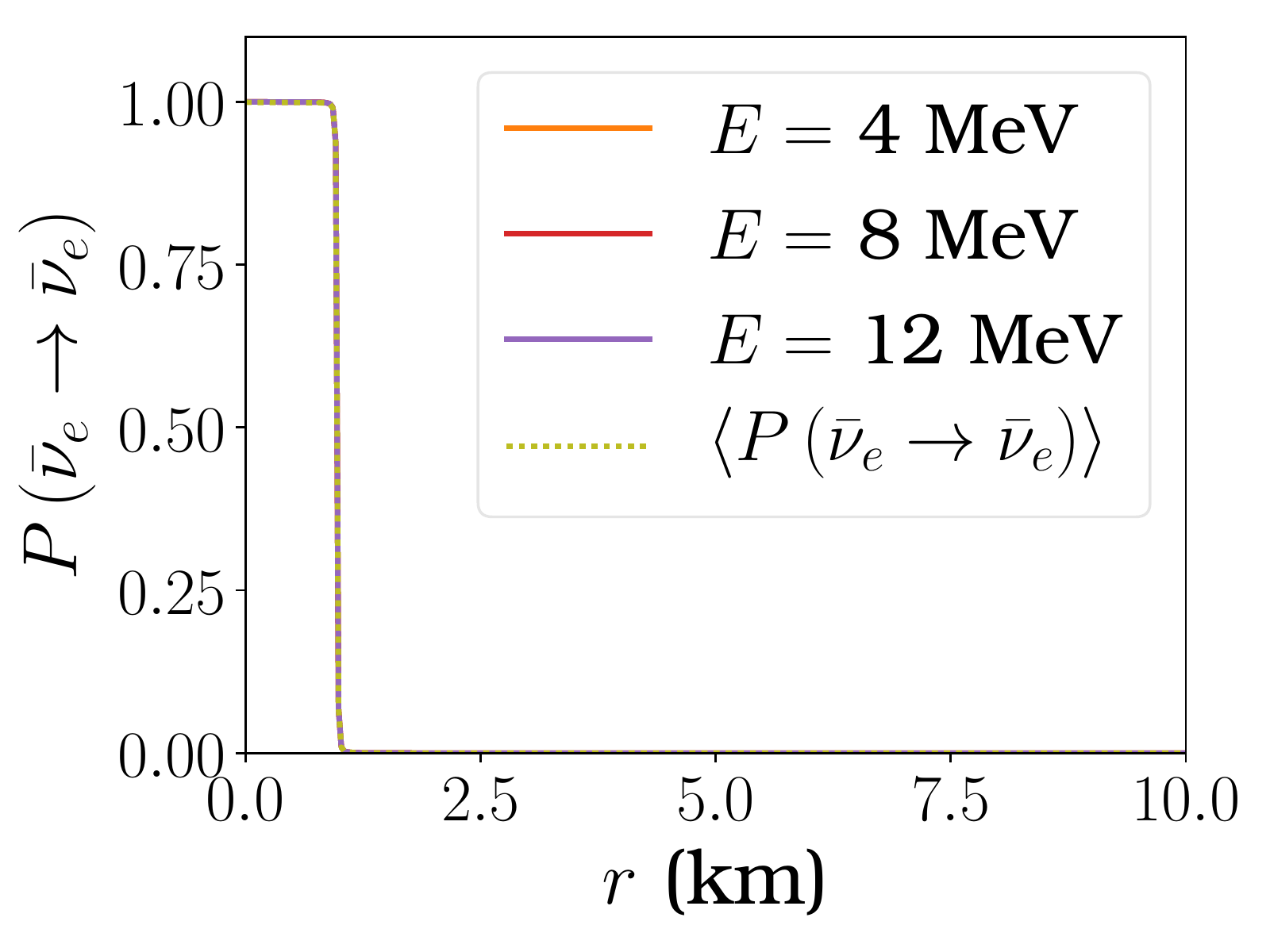}
\caption{Left panel : Difference of the diagonal elements of the total neutrino Hamiltonian (solid line), 
matter  potential  $V_M$ (dashed line) Eq.(\ref{e:Ires}) in presence of NSI contributions 
with $\delta \epsilon^n = -0.7$ and $\epsilon_0 = 1\times 10^{-4}$ and self-interaction oscillated potential (dotted line), as a function of distance from the emission point. The initial parameters are $x_0 =-30$ km, $z_0 =20$ km, and $\theta_q =55^\circ$. 
Middle and right panels :
Survival probabilities  for neutrinos (middle), antineutrinos (right).
Different energies corresponding to different colors as well as averaged probability (dotted line) are indistinguishable. 
The results are obtained by using baryon densities and electron fraction from the detailed simulations \cite{Perego:2014fma}.}
\label{fig:modC_70_e-4}
\end{figure*}

The I resonance occurs when the difference between the diagonal elements of the total Hamiltonian goes to zero, requiring for the total matter potential to meet the condition
\begin{multline}\label{e:Ires}
V_M  \equiv \lambda  \left[ Y_e  + \frac{ Y_\odot - Y_e }{Y_\odot} \delta \epsilon^n \right]  \\
\approx 2 \omega c_{2\theta} - \left( h^{\nu\nu}_{ee} -h^{\nu\nu}_{xx} \right).
\end{multline}
Refs.\cite{EstebanPretel:2007yu,EstebanPretel:2009is,Stapleford:2016jgz}  have pointed out that the presence of $\nu\nu$ self-interactions have negligible effects on the location and adiabaticity of the I resonance, thus making it to occur when the matter potential Eq.(\ref{e:Ires}) is very small. 

First we consider here a case in which the self-interaction potential is sub-dominant compared to the matter one.  In such cases the location of the I resonance coincides with the point where the matter potential $V_{\rm mat} $ becomes very small, which is possible in the presence of NSI because of a cancellation between the standard matter term and the nonstandard contribution. Figure \ref{fig:modC_70_e-4} (left panel) presents the difference of the diagonal elements, the total matter potential with $\delta \epsilon^n = -0.88$ and $\epsilon_0 = 1\times 10^{-4}$ and the oscillated self-interaction potential.
Condition (\ref{e:Ires}) can be satisfied for both neutrinos and antineutrinos simultaneously and is very little dependent on the neutrino energy.
Depending on the value of the diagonal NSI parameter $\delta \epsilon^n$, the I resonance can arise extremely close to the neutrinosphere, as already pointed out in the literature. In this example, it occurs at $1$ km from it. 

The survival probabilities for neutrinos and antineutrinos as well as the averaged one are shown in Figure \ref{fig:modC_70_e-4} for different neutrino energies (middle and right panels). 
Given a specific matter profile,  the resonance location only depends on the value of the diagonal NSI parameter, $\delta \epsilon^n$; whereas the value of $\epsilon_0$ impacts the adiabaticity. For the case shown,
the I resonance is adiabatic and induces significant conversion for both neutrinos and antineutrinos.
 It is worth noting that even in the presence of a small $\epsilon_0$ parameter, the flavor conversion behaviors stay independent of the energy. This is due to the fact that the off-diagonal self-interaction contribution to the Hamiltonian is, at the considered location, 
 much larger than the vacuum one, therefore suppressing the energy dependence. 

\subsubsection{I resonance as a synchronized MSW}
\begin{figure*}
\includegraphics[width=.3\textwidth]{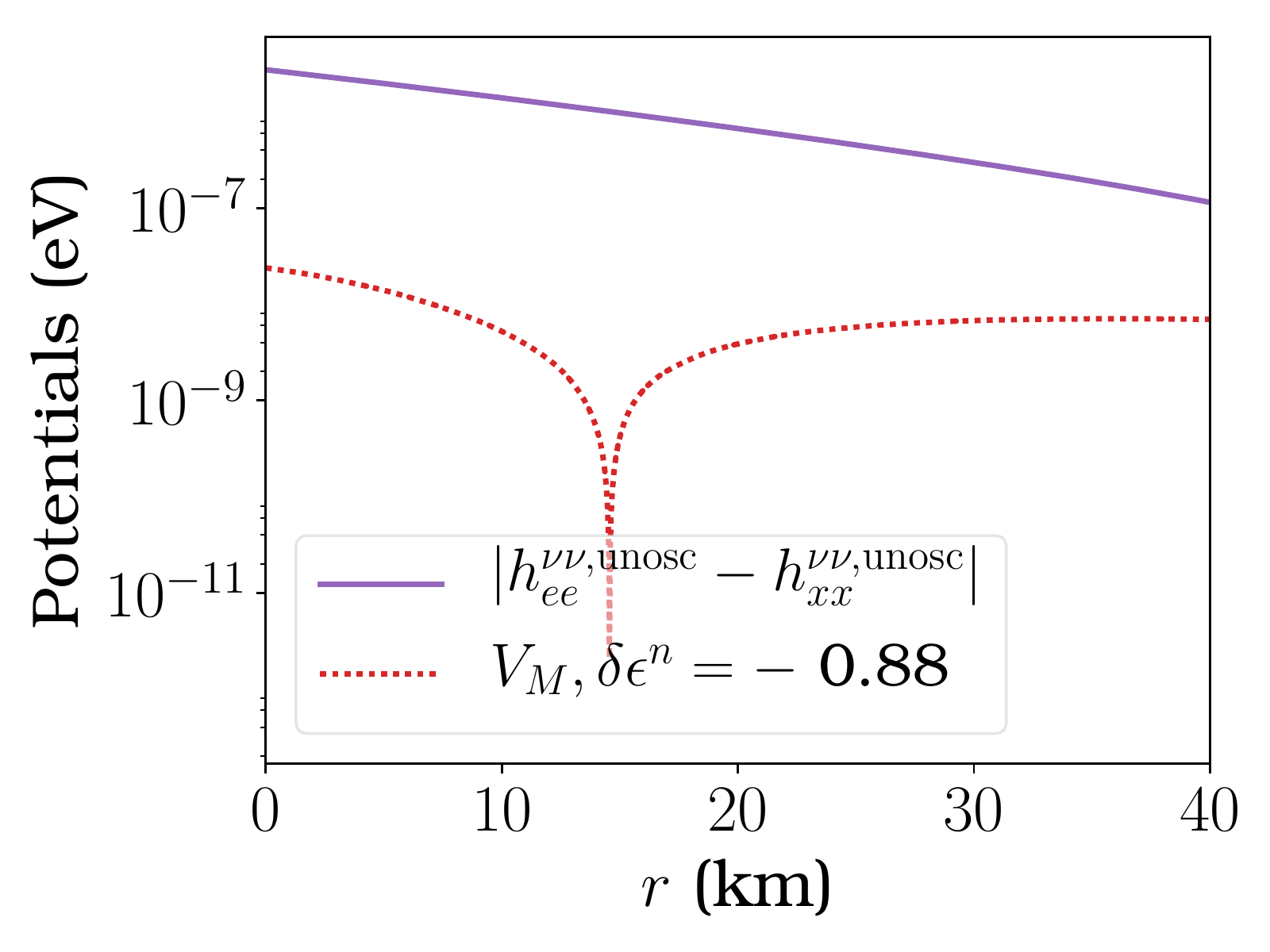}
\includegraphics[width=.3\textwidth]{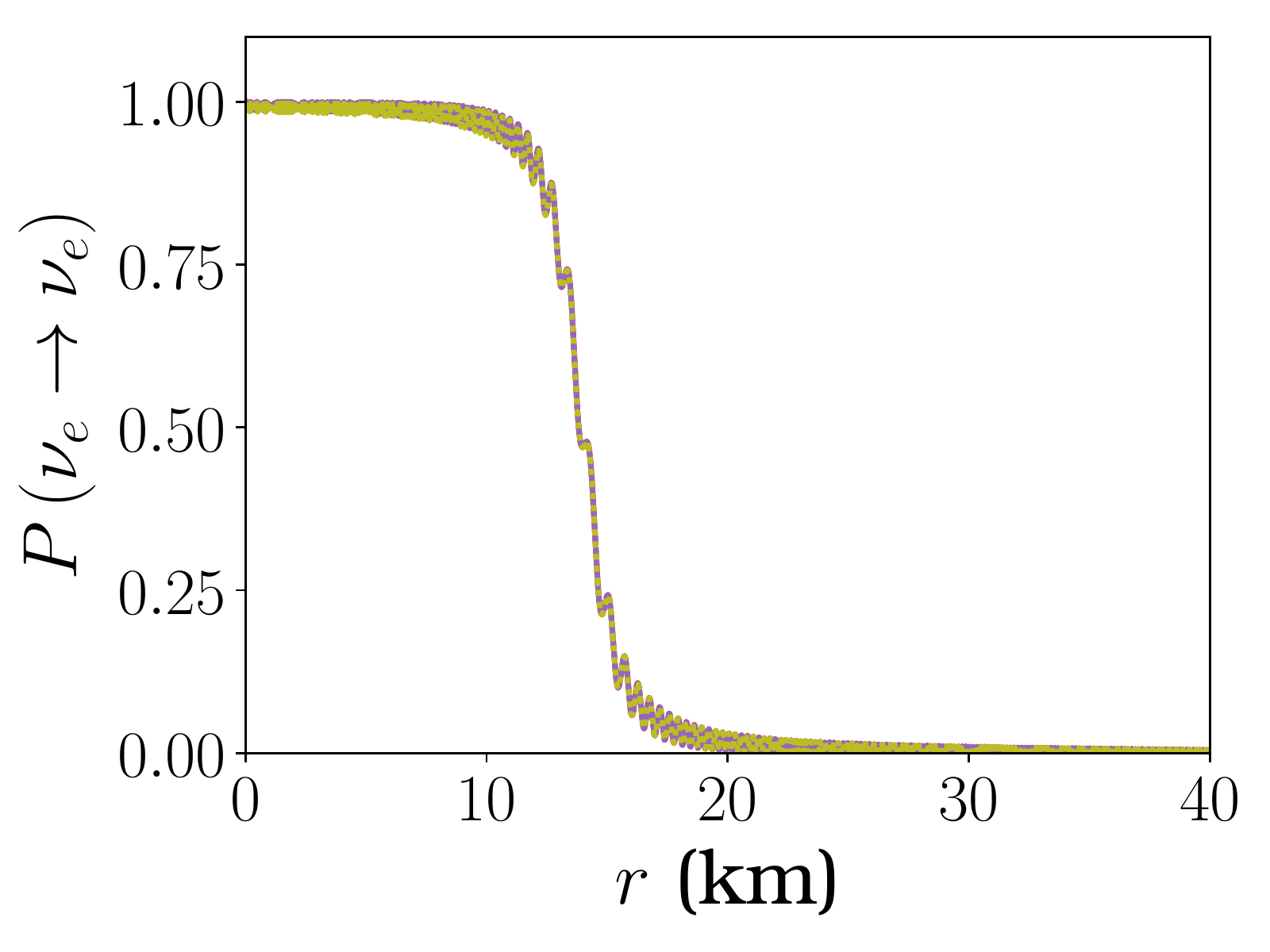}
\includegraphics[width=.3\textwidth]{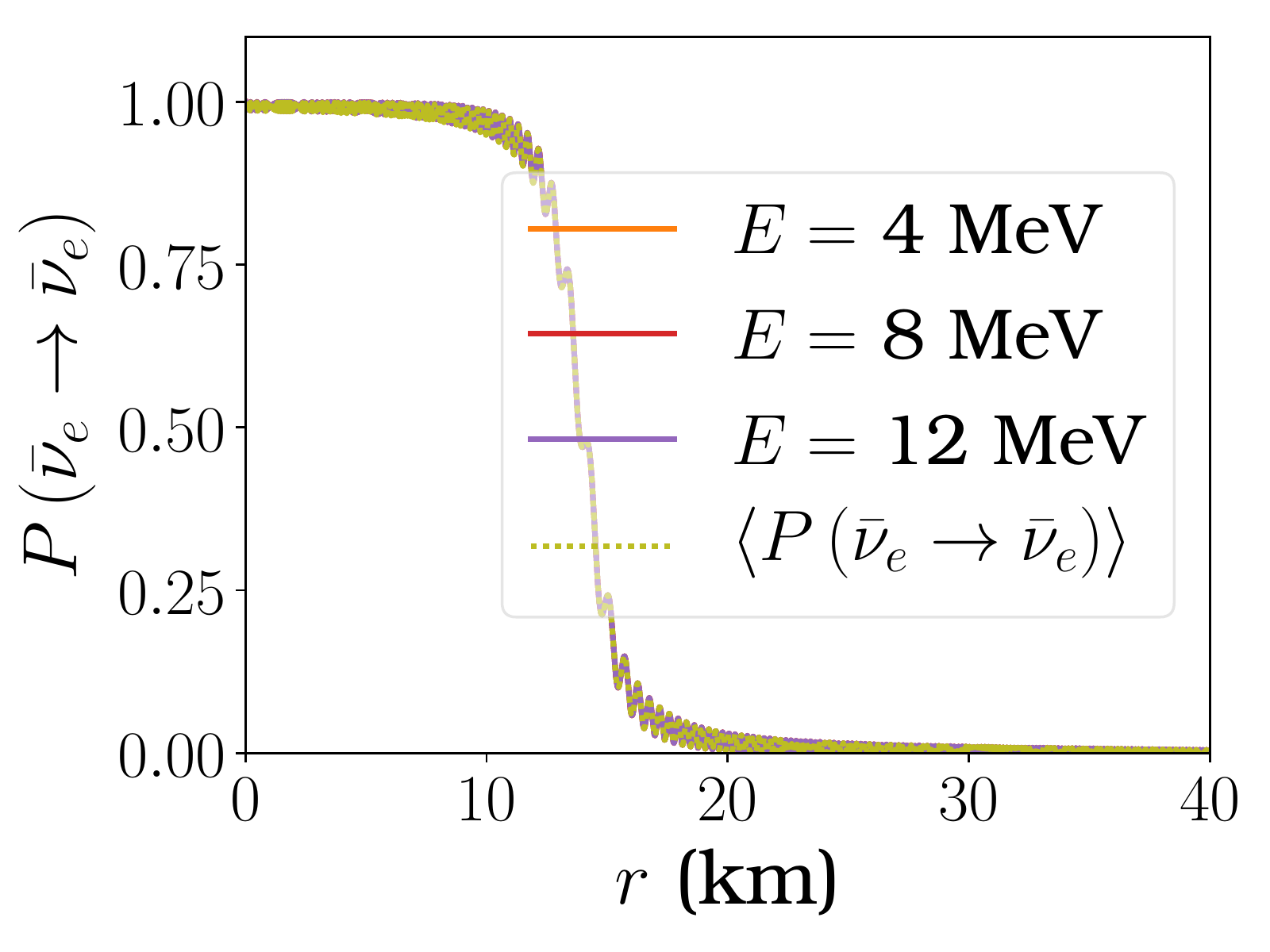}
\caption{Left panel : 
Matter  potential  $V_M$ (solid line) Eq.(\ref{e:Ires}) in presence of NSI contributions 
$\delta \epsilon^n = -0.88$ and $\epsilon_0 = 1\times 10^{-4}$ and self-interaction unoscillated potential Eq.(\ref{e:muunosc}) (dotted line), as a function of distance from the emission point. The initial parameters are $x_0 = 15$ km, $z_0 = 32$ km, and $\theta_q = 15^\circ$. 
Middle and right panels :
Survival probabilities for neutrinos (middle), antineutrinos (right).
Different energies corresponding to different colors as well as averaged probability (dotted line) are indistinguishable.}
\label{fig:modA_88_e-4}
\end{figure*}

While exploring the parameter space and different trajectories for the neutrino propagation, we have encountered situations where, although the self-interaction unoscillated potential is several orders of magnitude larger than the matter potential, a I resonance takes place and leads to significant flavor conversions. 
Figure \ref{fig:modA_88_e-4} shows a typical example of this situation with the NSI parameters $\delta \epsilon^n = -0.88$ and $\epsilon_0 = 1\times 10^{-4}$.
One can see that although
the unoscillated self-interaction potential $\mu$ (\ref{e:muunosc}) dominates the matter one $\lambda$ (\ref{e:hmat2f}), flavor conversions occur at the same location where the I resonance condition is fulfilled. Note that the difference between the self-interaction oscillated diagonal elements do cancel at the same point. 
We will be unraveling this effect in the light of synchronized flavor conversions.

\paragraph{Spin description}
In order to describe this phenomenon, we use the $\mathrm{SU} \lp 2 \rp$  isospin formalism in flavor space. The effective isospin vector $\vec{P}_{\nu_{\underline{\alpha}}} \lp r, q \rp$ denoting a neutrino of initial flavor $\alpha$ is related to the neutrino density matrix according to
\begin{equation}\label{e:P}
\rho_{\nu_{\underline{\alpha}}} \lp r, q \rp = \frac{1}{2} \lp  \mathbb{I} + \vec{\sigma} \cdot \vec{P}_{\nu_{\underline{\alpha}}} \lp r, q \rp \rp,
\end{equation}
and similarly for antineutrinos, where $\mathbb{I}$ is the $2\times2$ identity matrix and $\vec{\sigma} = \lp \sigma_x, \sigma_y, \sigma_z \rp$ is a vector in flavor space whose components are the Pauli $\sigma$ matrices. In this theoretical framework, the Liouville Von-Neumann equations are replaced by precession equations for $\vec{P}_{\nu_{\underline{\alpha}}} \lp r, q \rp$ with an effective magnetic field defined as 
\begin{equation}
h \lp r, q \rp = \frac{1}{2} \lp  \mathbb{I} + \vec{\sigma} \cdot \vec{B} \lp r, q \rp \rp.
\label{eq:Beff}
\end{equation}
and receiving three contributions
\begin{equation}\label{e:B}
\vec{B} \lp r, q \rp = \vec{B}_\text{vac} \lp q \rp + \vec{B}_\text{mat} \lp r \rp   + \vec{B}_{\nu\nu} \lp r \rp,
\end{equation}
Note that the expressions for $\vec{\bar{P}}_{\bar{\nu}_{\underline{\alpha}}}$ and
$\vec{\bar{B}}$ are analogous to Eqs.(\ref{e:P}) and (\ref{e:B}) respectively. 
In the antineutrino case, the vacuum contribution in Eq.(\ref{e:B}) has a minus sign.
The vacuum term is given by
\begin{equation}
\vec{B}_\text{vac} = 2 \omega \vec{B}_0 = 2 \omega \begin{pmatrix}
s_{2\theta} \\ 0 \\ -c_{2\theta}, 
\end{pmatrix}
\end{equation}
 while the matter term includes the standard and nonstandard contributions
 \begin{equation}
\vec{B}_\text{mat} =  \lambda \left[ Y_e \begin{pmatrix}
0 \\ 0 \\ 1
\end{pmatrix} + \begin{pmatrix}
2 \lp 3 + Y_e  \rp \mathrm{Re} \epsilon_0 \\ - 2 \lp 3 + Y_e  \rp \mathrm{Im} \epsilon_0 \\ \delta \epsilon^n \lp \frac{Y_\odot - Y_e}{Y_\odot} \rp .
\end{pmatrix} \right]
\end{equation}
The third term in Eq.(\ref{e:B}) comes from the self-interaction term of the neutrino Hamiltonian
\begin{multline}
\vec{B}_{\nu\nu}   = \sqrt{2} G_F \sum_{\alpha = e,x} \int_0^\infty \! \mathrm dp \lp G_{\nu_\alpha}  j_{\nu_\alpha} \lp p \rp \vec{P}_{\nu_{\underline{\alpha}}} \lp  p \rp \right.\\
\left. - G_{\bar{\nu}_\alpha} j_{\bar{\nu}_\alpha} \lp p \rp \vec{\bar{P}}_{\bar{\nu}_{\underline{\alpha}}} \lp p \rp \rp,
\end{multline}
where $j_{\nu_\alpha} \lp p \rp  = \frac{L_{\nu_\alpha} f_{\nu_\alpha} \lp p \rp }{\pi^2 R_{\nu_\alpha}^2 \left< E_{\nu_\alpha} \right>}$ and similarly for antineutrinos. Note that the explicit r-dependences are not shown for readability.

In order to describe the collective neutrino mode associated to the I resonance let us introduce the $\vec{J}$ vector 
\begin{multline}\label{e:Jvec}
\vec{J}  = \sum_{\alpha = e, x} \int_0^\infty \! \mathrm dp \lp G_{\nu_\alpha}  j_{\nu_\alpha} \lp p \rp \vec{P}_{\nu_{\underline{\alpha}}} \lp p \rp \right.\\
\left. - G_{\bar{\nu}_\alpha}  j_{\bar{\nu}_\alpha} \lp p \rp \vec{\bar{P}}_{\bar{\nu}_{\underline{\alpha}}} \lp p \rp \rp.
\end{multline}
We emphasize that,  in a BNS merger scenario, one needs to include the geometrical factors in the definition of the collective vector,
contrarily to what is usually done  in the bulb model for supernovae (single-angle approximation), as e.g. in \cite{Pastor:2001iu}. The reason is that here the geometrical factors differ for different flavors even when one employs the {\it ansatz} given by Eq.(\ref{e:st}). 
With definition (\ref{e:Jvec}) one can write the neutrino self-interaction term proportional to a unique vector $\vec{J}$, namely
\beq\label{e:Jvec2}
\vec{B}_\text{self}  = \sqrt{2} G_F \vec{J}.
\eeq

The evolution equation for $\vec{J}$ can be derived from the ones of $\vec{P}_{\nu_{\underline{\alpha}}}$ (and $\vec{\bar{P}}_{\bar{\nu}_{\underline{\alpha}}}$) and using the explicit expressions of $\vec{B}$ ($\vec{\bar{B}}$). One finds 
\begin{widetext}
\begin{multline}
\partial_r \vec{J} = \vec{B}_\text{mat} \times \vec{J} 
+ \vec{B}_0 \times \sum_{\alpha = e,x} \int_0^\infty \! \mathrm dp \frac{\Delta m^2}{2p} \lp G_{\nu_\alpha} j_{\nu_\alpha} \lp p \rp \vec{P}_{\nu_{\underline{\alpha}}} \lp p \rp  +\ G_{\bar{\nu}_\alpha} j_{\bar{\nu}_\alpha} \lp p \rp \vec{\bar{P}}_{\bar{\nu}_{\underline{\alpha}}} \lp p \rp \rp \\ 
+  \sum_{\alpha = e,x} \int_0^\infty \! \mathrm dp \lp \partial_r G_{\nu_\alpha} j_{\nu_\alpha} \lp p \rp \vec{P}_{\nu_{\underline{\alpha}}} \lp p \rp -\ \partial_r G_{\bar{\nu}_\alpha}  j_{\bar{\nu}_\alpha} \lp p \rp \vec{\bar{P}}_{\bar{\nu}_{\underline{\alpha}}} \lp p \rp \rp.
\end{multline}
\end{widetext}

Let us assume now that, during the evolution, the modes all start along the z-axis, i.e. $\vec{P}_{\nu_\alpha} \lp r, p \rp \approx P_{\nu_\alpha, z } \lp 0, p \rp \hat{J}$ and stay aligned with the collective mode $\vec{J}$ (similarly for antineutrinos). 
If neutrinos and antineutrinos of any momentum stay synchronized in flavor space during the propagation, the evolution equation for $\vec{J}$ becomes

\begin{widetext}
\begin{equation}\label{e:Jeq}
\partial_r \vec{J} \approx \vec{B}_\text{mat} \times \vec{J} \\
+ \vec{B}_0 \times \hat{J} \int_0^\infty \! \mathrm dp \frac{\Delta m^2}{2p} \left[ G_{\nu_e} j_{\nu_e} \lp p \rp  +\ G_{\bar{\nu}_e} j_{\bar{\nu}_e} \lp p \rp   - 2 G_{\nu_x} j_{\nu_x} \lp p \rp \right] + \hat{J} \frac{\partial_r \mu}{\sqrt{2} G_F}.
\end{equation} 
\end{widetext}
While the first two terms are ordinary oscillation terms, the last one is a damping term, taking into account that the norm of this collective mode decreases with time. This is due to the fact that the geometry of the problem is included in the definition of $\vec{J}$. Note that 
such a decrease should not be interpreted as lepton number conservation violation, but as a neutrino density decrease  along a given trajectory, due to the geometry. Let us characterize this decrease by multiplying the evolution equation (\ref{e:Jeq}) by $\vec{J}$ 
\begin{equation}
\vec{J} \cdot \partial_r \vec{J} = \frac{1}{2} \partial_r \vec{J}^2 \approx \left| \vec{J} \right| \frac{\partial_r \mu}{\sqrt{2} G_F},
\end{equation}
which gives $\left| \vec{J} \lp r \rp \right| \approx \frac{\mu \lp r \rp}{\sqrt{2} G_F}$. Plugging this  expression in Eq.(\ref{e:Jeq}), one finds 

\begin{equation}
\partial_r \vec{J} \approx \vec{B}_J \times \vec{J} + \hat{J} \frac{\partial_r \mu}{\sqrt{2} G_F}.
\end{equation} 
The effective magnetic field associated with he collective mode $\vec{J}$ is $\vec{B}_J = \omega_\text{sync} \vec{B}_0 + \vec{B}_\text{mat}$ 
which components are
\begin{equation}
\vec{B}_J = \begin{pmatrix}
2 \lambda \lp 3+Y_e \rp \mathrm{Re} \epsilon_0 + \omega_\text{sync} s_{2\theta}  \\
-2 \lambda \lp 3+Y_e \rp \mathrm{Im} \epsilon_0  \\
-\omega_\text{sync} c_{2\theta} + V_M
\end{pmatrix}.
\label{eq:bjcomp}
\end{equation} 
The synchronized frequency  $\omega_\text{sync}$ is  $\vec{J}$ precession frequency
\begin{multline}
\omega_\text{sync}  = \frac{\sqrt{2} G_F}{\mu} \int_0^\infty \! \mathrm dp \frac{\Delta m^2}{2p} \left[ G_{\nu_e}  j_{\nu_e} \lp p \rp   \right. \\
\left. +\ G_{\bar{\nu}_e}  j_{\bar{\nu}_e} \lp p \rp   - 2 G_{\nu_x} j_{\nu_x} \lp p \rp \right].
\end{multline}

Assuming the fluxes follow Fermi-Dirac distributions, the integral above can be computed, and $\omega_\text{sync}$ can be expressed as 
\begin{widetext}
\begin{equation}
\omega_\text{sync}    = \frac{\sqrt{2} G_F \Delta m^2 F_1 \lp 0 \rp F_3 \lp 0 \rp}{2 \mu   F^2_2 \lp 0 \rp} \left[ \frac{L_{\nu_e} G_{\nu_e}}{ R^2_{\nu_e} \left\langle E_{\nu_e}\right\rangle^2 }  + \frac{L_{\bar{\nu}_e} G_{\bar{\nu}_e}}{ R^2_{\bar{\nu}_e} \left\langle E_{\bar{\nu}_e}\right\rangle^2 } - 2 \frac{L_{\nu_x} G_{\nu_x}}{ R^2_{\nu_x} \left\langle E_{\nu_x}\right\rangle^2 } \right].
\label{eq:omegasync}
\end{equation}
\end{widetext}

\paragraph{Resonance condition} In addition to a precession motion, the collective mode $\vec{J}$ can also meet a MSW-like resonance condition $B_{J,z} \approx 0$, which requires
\begin{equation}
\omega_\text{sync} \lp r_I \rp  c_{2\theta} = V_M \lp r_I \rp,
\label{eq:rescondsync}
\end{equation}
where $r_I$ is the resonance location. From \Eqn{eq:omegasync}, it can be seen that $\omega_\text{sync} \propto \frac{1}{\mu}$ : in situations where the neutrino background dominates, the {\it l.h.s.} of \Eqn{eq:rescondsync} is often several of magnitude smaller than the {\it r.h.s.}. However, in cases where the total matter potential $V_M$ goes to zero, this resonance condition can be met. The reversed situation, in which the resonance condition is met because $\mu$ goes to zero, has been already pointed out in \cite{Frensel:2016fge}.

\begin{figure}[h!]
\includegraphics[width=.3\textwidth]{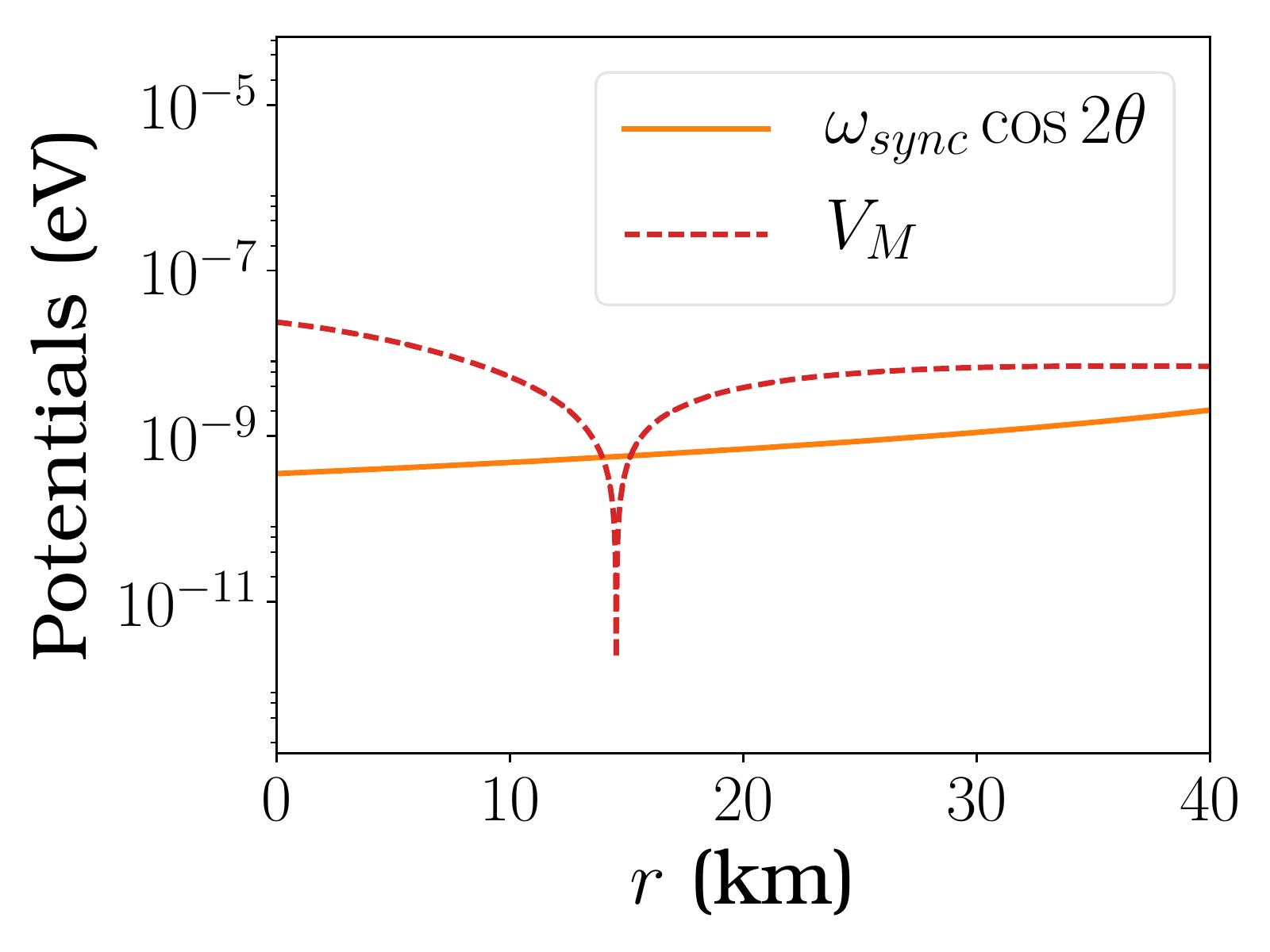}
\caption{Contributions to the $z$ component of the effective magnetic field $\vec{B}_J$. The solid line represents $\omega_\text{sync} c_{2\theta}$, while the dashed line shows the matter potential $V_M$. It can be seen that when $V_M$ cancels, due to the presence of NSI, the synchronized resonance condition \Eqn{eq:rescondsync} is met. The NSI and trajectory parameters used here are the same as the ones used in figure \ref{fig:modA_88_e-4}.}
\label{fig:syncrescond}
\end{figure}

\begin{figure*}
\includegraphics[width=.3\textwidth]{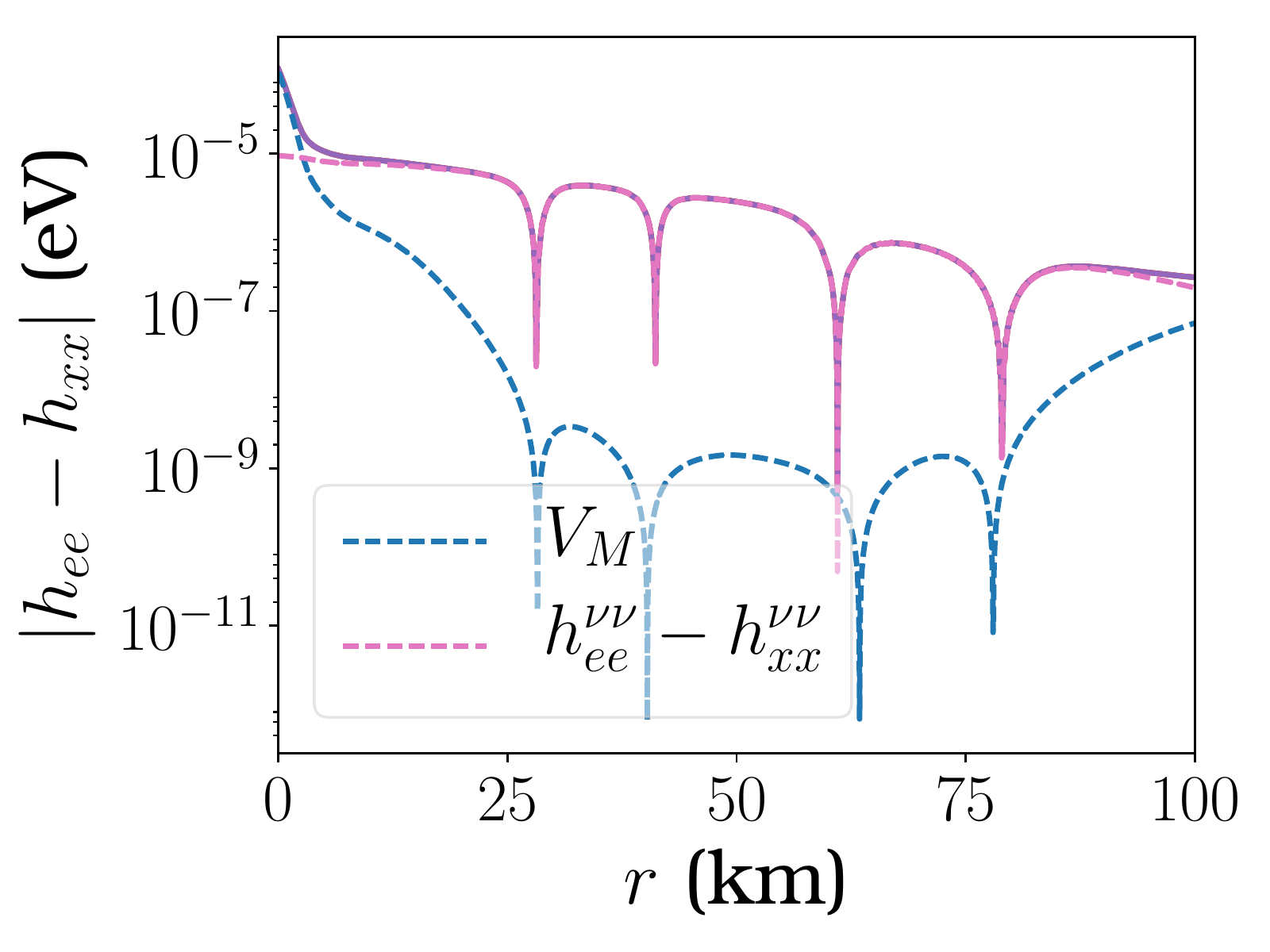}
\includegraphics[width=.3\textwidth]{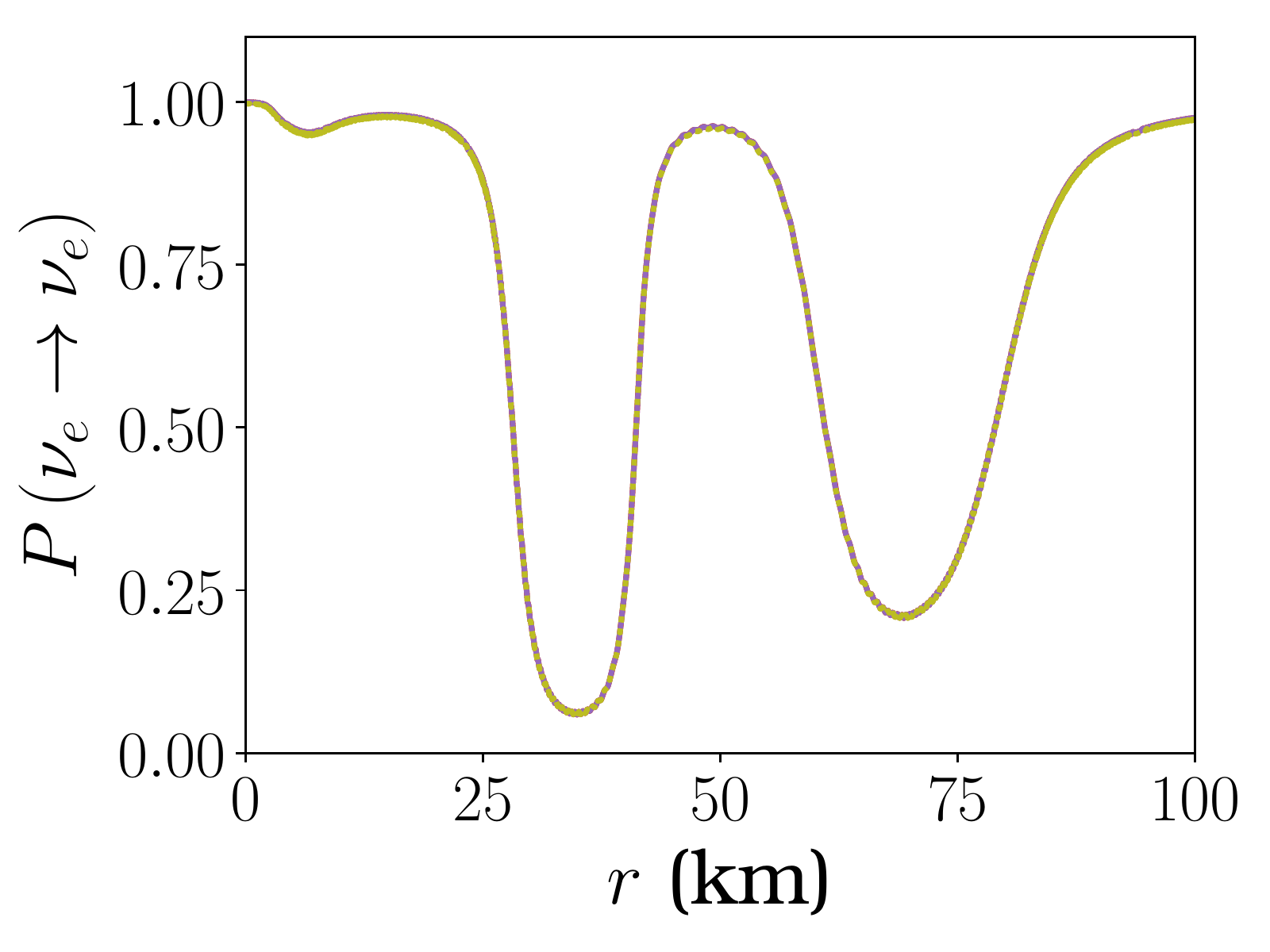}
\includegraphics[width=.3\textwidth]{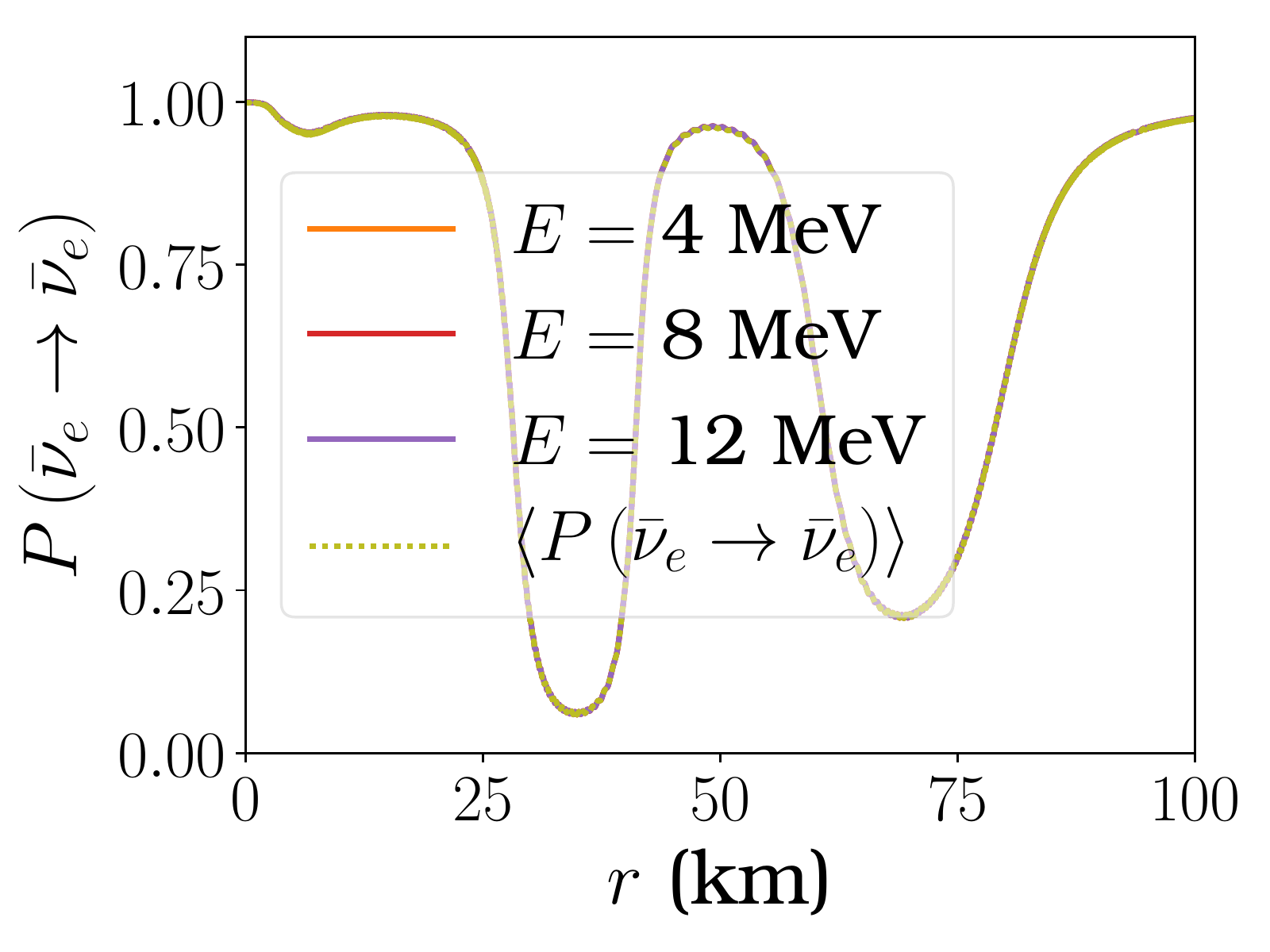}
\caption{Left panel : Difference of the diagonal elements of the total neutrino Hamiltonian (solid line), matter  potential  $V_M$ (dashed line) Eq.(\ref{e:Ires}) in presence of NSI contributions with $\delta \epsilon^n = -0.90$ and $\epsilon_0 = 1\times 10^{-3}$ and self-interaction oscillated potential (dotted line), as a function of distance from the emission point. The initial parameters are $x_0 =-35$ km, $z_0 =25$ km, and $\theta_q =50^\circ$.  Middle and right panels : Survival probabilities  for neutrinos (middle), antineutrinos (right). Different energies corresponding to different colors as well as averaged probability (dotted line) are indistinguishable. Several synchronized I resonances are present in this case, at $29$ km, $40$ km, $65$ km and $78$ km.}
\label{fig:modD_90_e-3}
\end{figure*}

Figure \ref{fig:syncrescond} shows the {\it r.h.s.} and the  {\it l.h.s.} of \Eqn{eq:rescondsync} corresponding to the case of Figure \ref{fig:modA_88_e-4}.
One can see that the synchronized MSW resonance condition given by Eq.(\ref{eq:rescondsync}) is met almost at location where $V_M$ goes to zero, i.e. at the location of the I resonance, as can be seen from the conversion probabilities.
Another example of synchronized I resonance is shown in Figure \ref{fig:modD_90_e-3} with the neutrino self-interaction dominating over the matter potential. Significant conversion can be seen at $29$  km, $40$  km, $65$  km et $78$ km.

\paragraph{Adiabaticity and influence of $\epsilon_0$}
In order to characterize further flavor conversion at the I resonance, we can define an adiabaticity parameter as 
\begin{equation}
\gamma = \left. \frac{| \vec{B}_J |^3}{| \frac{\mathrm d\vec{B}_J}{\mathrm dt}\! \times\! \vec{B}_J |} \right|_{r = r_I}.
\label{eq:adiabI}
\end{equation}
From \Eqn{eq:rescondsync}, it can be seen that the value of $\epsilon_0$ has no influence on the resonance location whereas  it influences the adiabaticity of the transformation.

%

\begin{figure*}
\includegraphics[width=.3\textwidth]{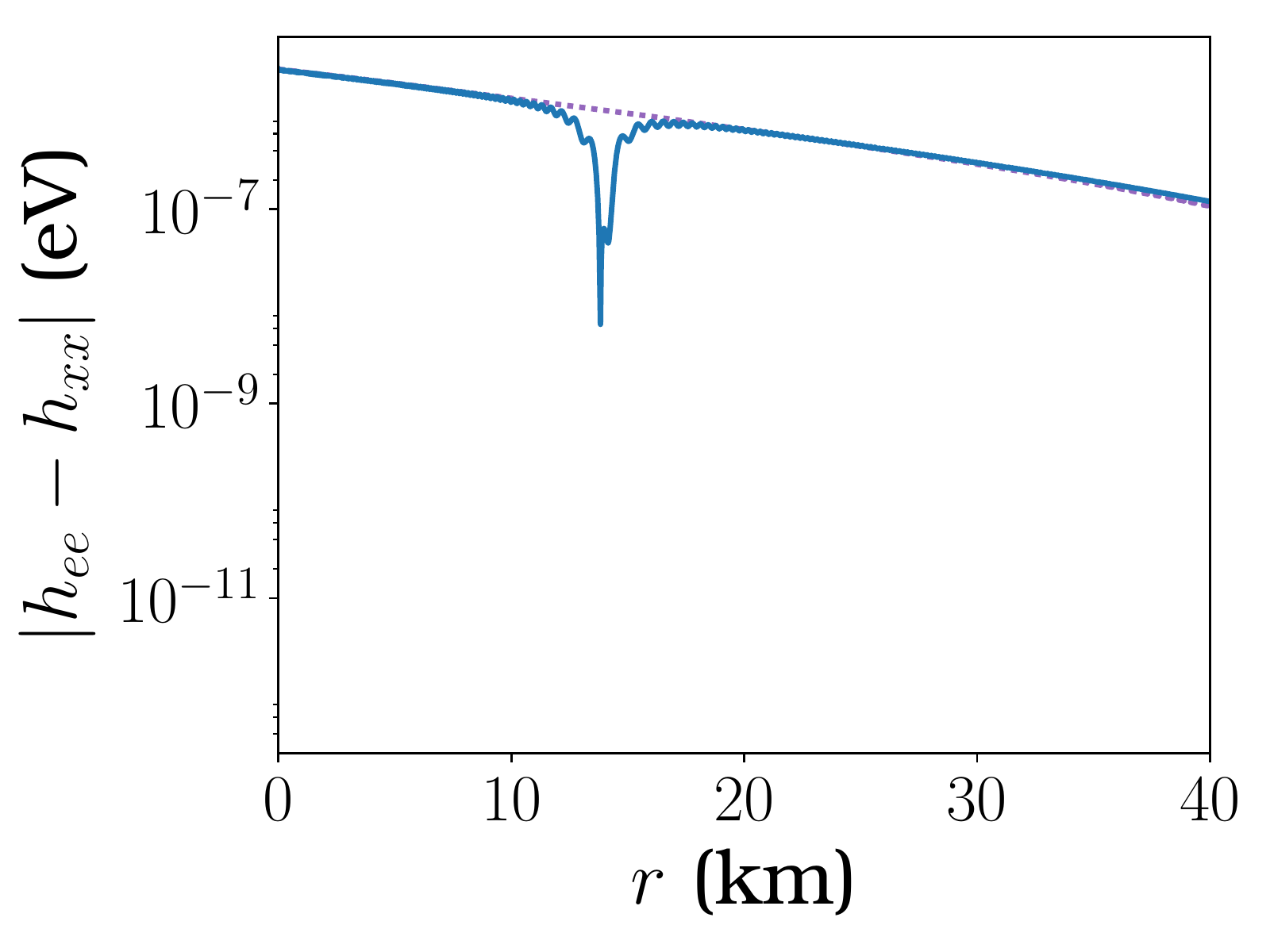}
\includegraphics[width=.3\textwidth]{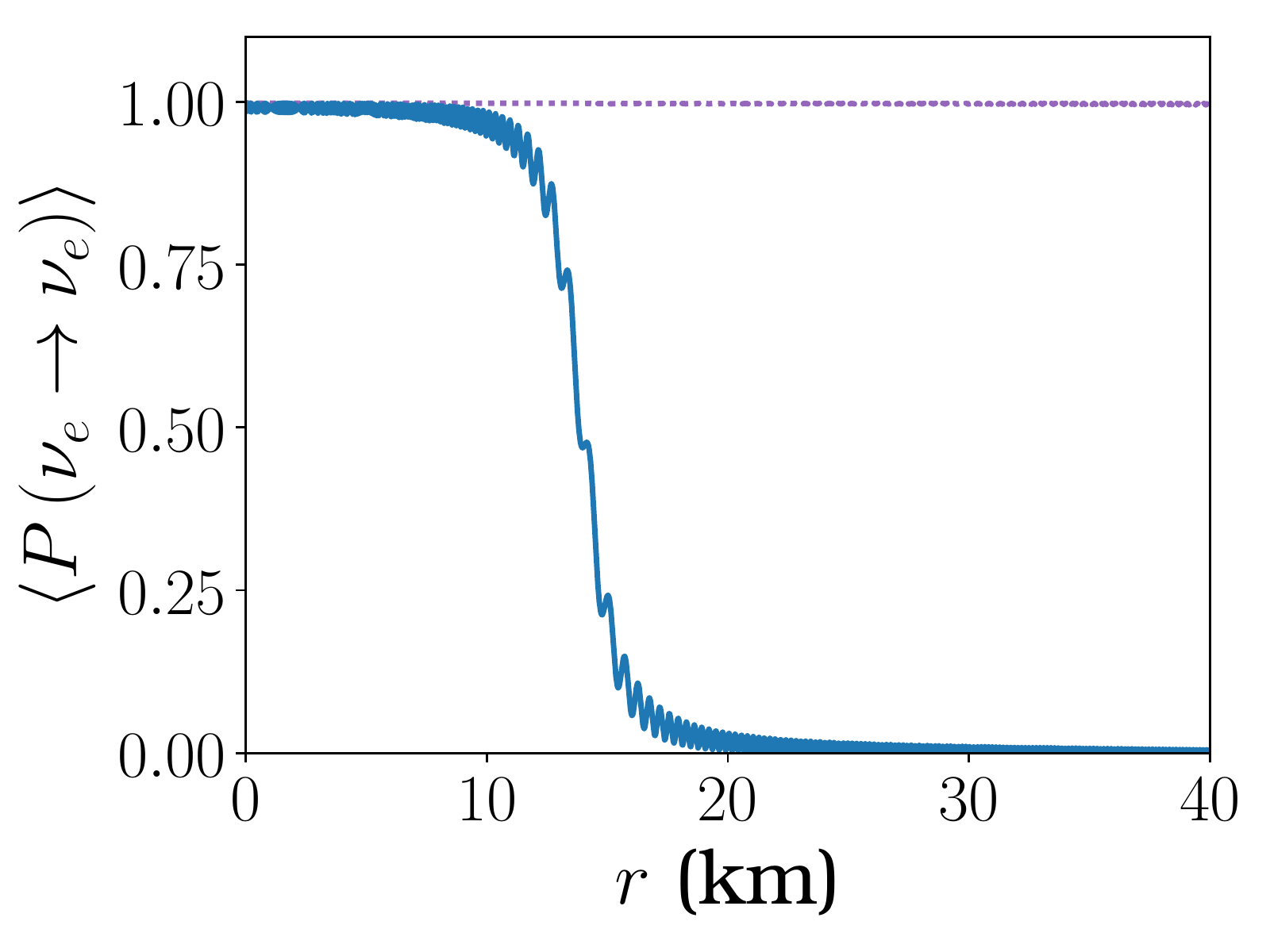}
\includegraphics[width=.3\textwidth]{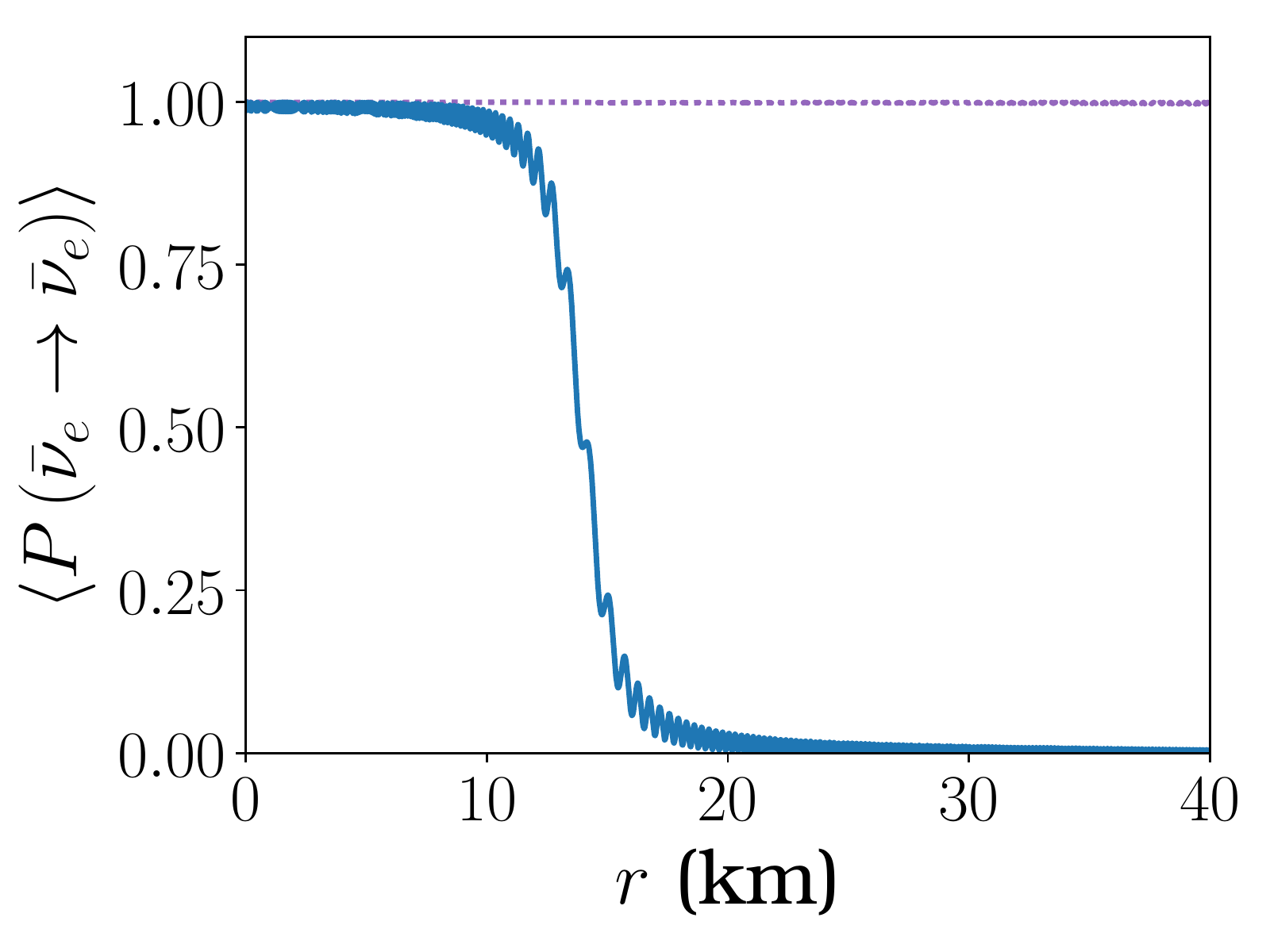}
\caption{Left panel : Difference of the diagonal elements of the total neutrino Hamiltonian, as a function of distance from the emission point. The initial parameters are $x_0 =15$ km, $z_0 =32$ km, and $\theta_q =15^\circ$. 
Middle and right panels :
Averaged survival probabilities  for neutrinos (middle), antineutrinos (right).  The NSI parameters are set to $\delta \epsilon^n = -0.88$ and $\epsilon_0 = 1\times 10^{-4}$ (solid lines) and $\epsilon_0 = 1\times 10^{-5}$ (dotted lines).}
\label{fig:modA_88_4vs5}
\end{figure*}

\begin{figure}
\includegraphics[width=.3\textwidth]{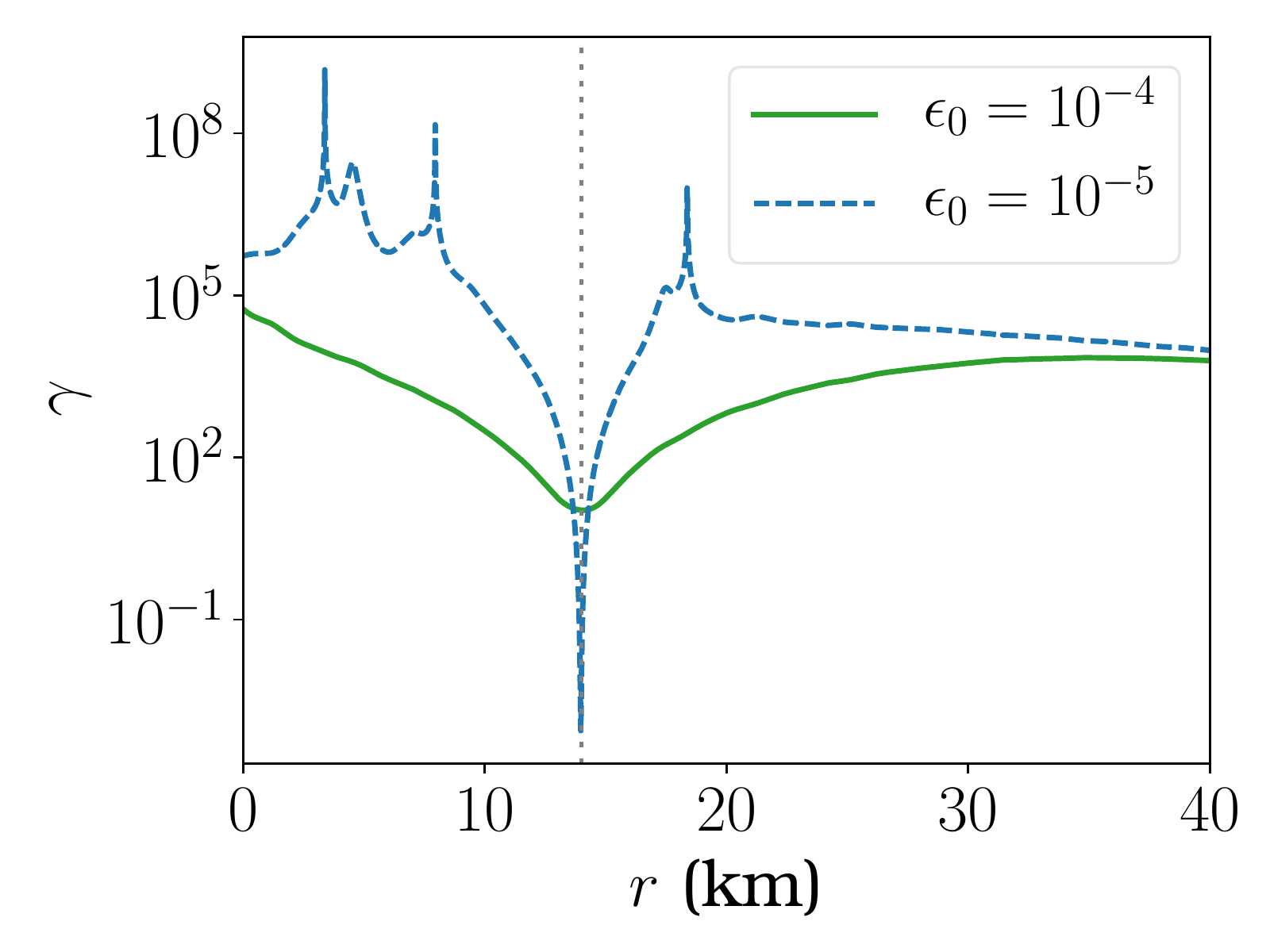}
\caption{Adiabaticity parameter as the right-hand-side of Eq.(\ref{eq:adiabI}), corresponding to Figure \ref{fig:modA_88_4vs5}. The solid line corresponds to $\epsilon_0 = 1\times 10^{-4}$, while the dashed line corresponds to $\epsilon_0 = 1\times 10^{-5}$. The location of the I resonance is shown as a vertical dotted line.}
\label{fig:adiab}
\end{figure}

Figure \ref{fig:modA_88_4vs5} shows an example of the influence of $\epsilon_0$ on the adiabaticity. Going from $\epsilon_0 = 10^{-4}$ to $ 10^{-5}$
the oscillation probabilities for neutrinos and antineutrinos go from complete flavor conversion from $\nu_e$ to $\nu_x$ to no conversion.
The adiabaticity parameter Eq.(\ref{eq:adiabI}) corresponding to this case is presented in Figure \ref{fig:adiab}. It can be seen that, at the location of the resonance, the adiabaticity parameter in the case of $\epsilon_0 = 1\times 10^{-5}$ is two order of magnitude smaller than the one for $\epsilon_0 = 1\times 10^{-4}$, consistently with the behaviors observed for the survival probabilities. Note that the cancellation of the adiabaticity parameter around the resonance in the case of $\epsilon_0 = 1\times 10^{-5}$ comes from the fact that for this value of $\epsilon_0$, the matter contribution and the $\omega_\mathrm{sync}$ contribution in $B_{J,x}$ (Eq.(\ref{eq:bjcomp})) are of the same order of magnitude and of opposite signs, making $B_{J,x}$ very small. Therefore, at the resonance, as $B_{J,z}$ tends to 0 , $\gamma \rightarrow  \frac{B^2_{J,x}}{\partial_r B_{J,z}}$ becomes much smaller at the same time. 

\begin{figure*}
\includegraphics[width=.3\textwidth]{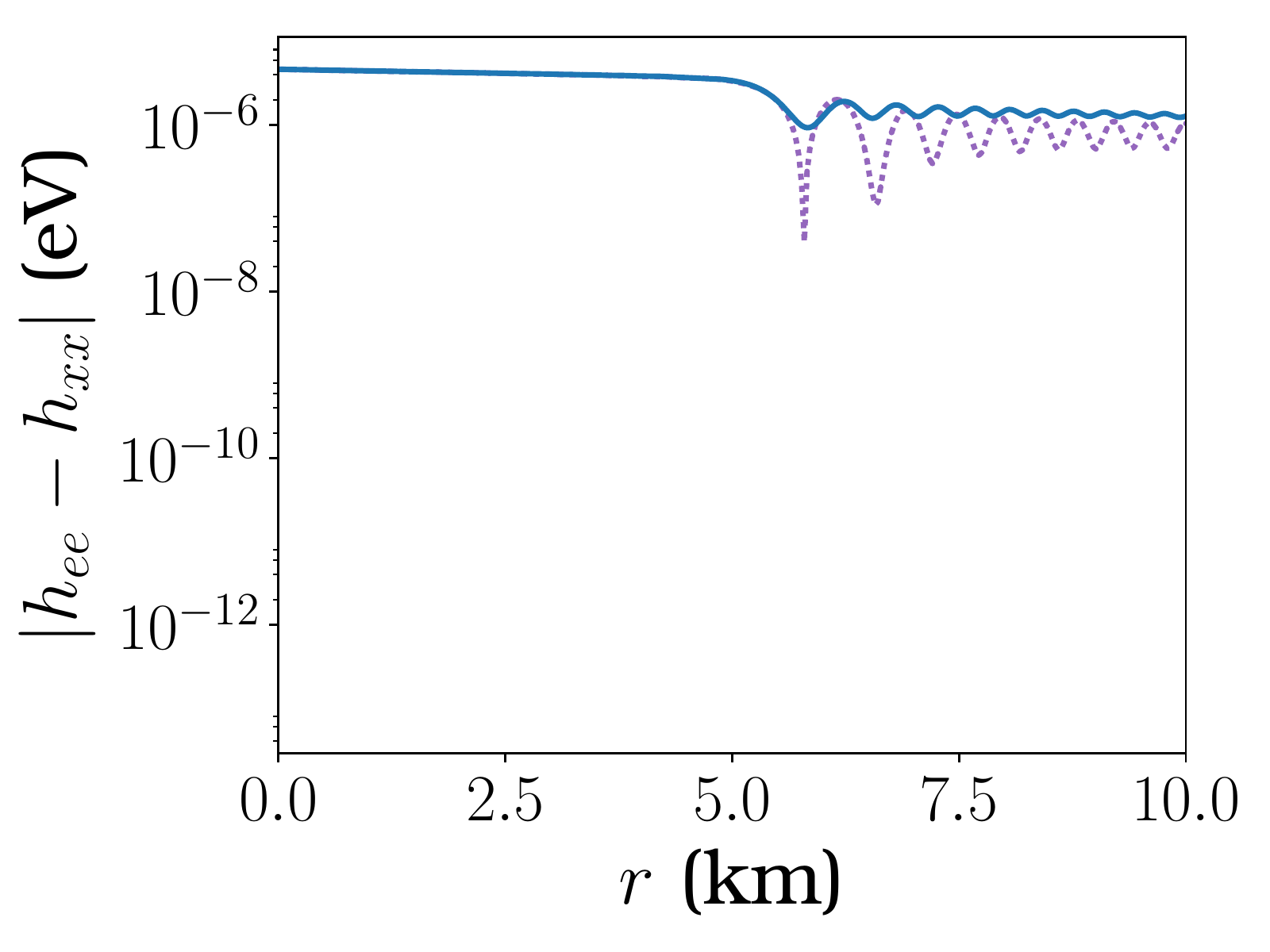}
\includegraphics[width=.3\textwidth]{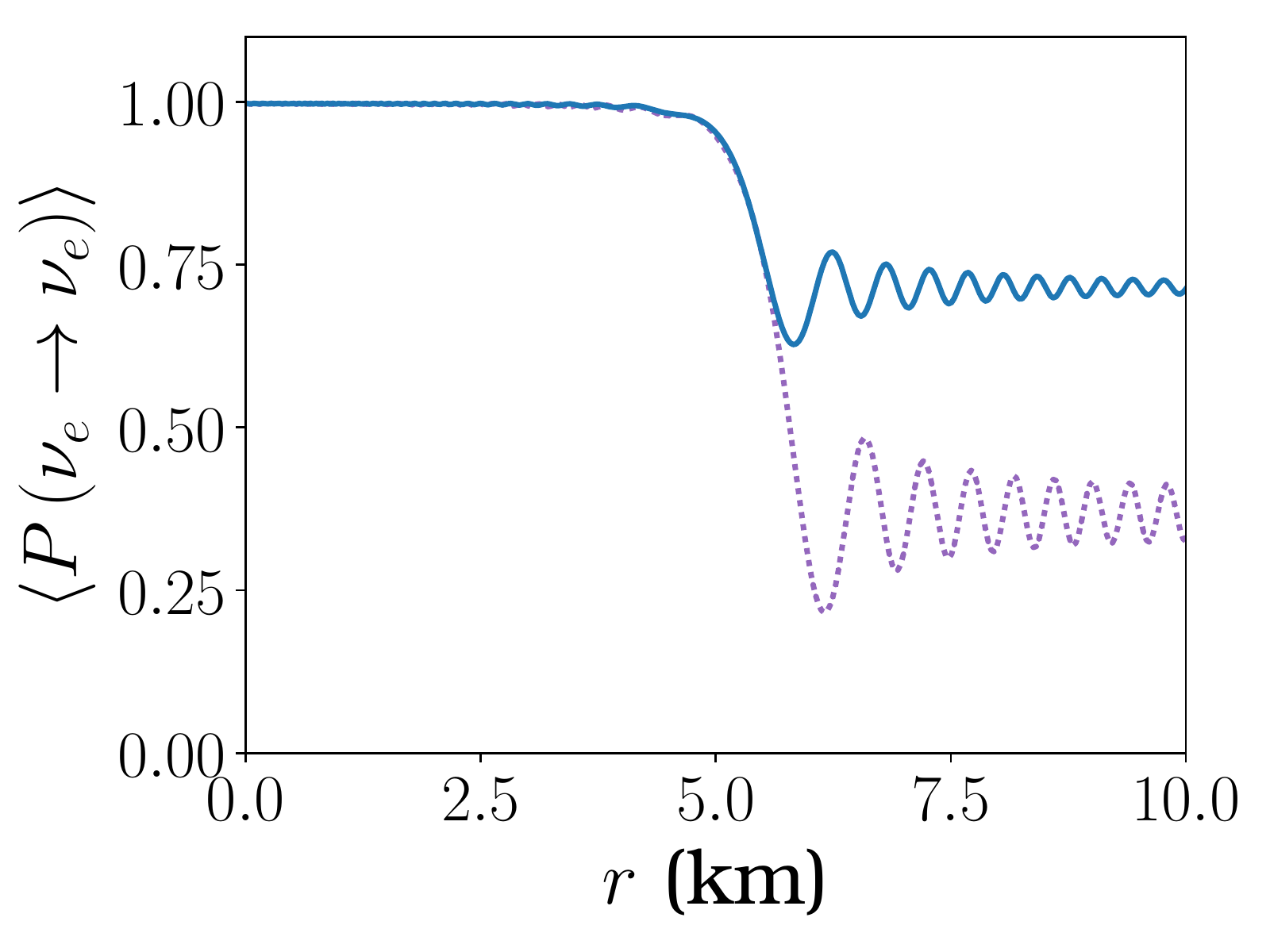}
\includegraphics[width=.3\textwidth]{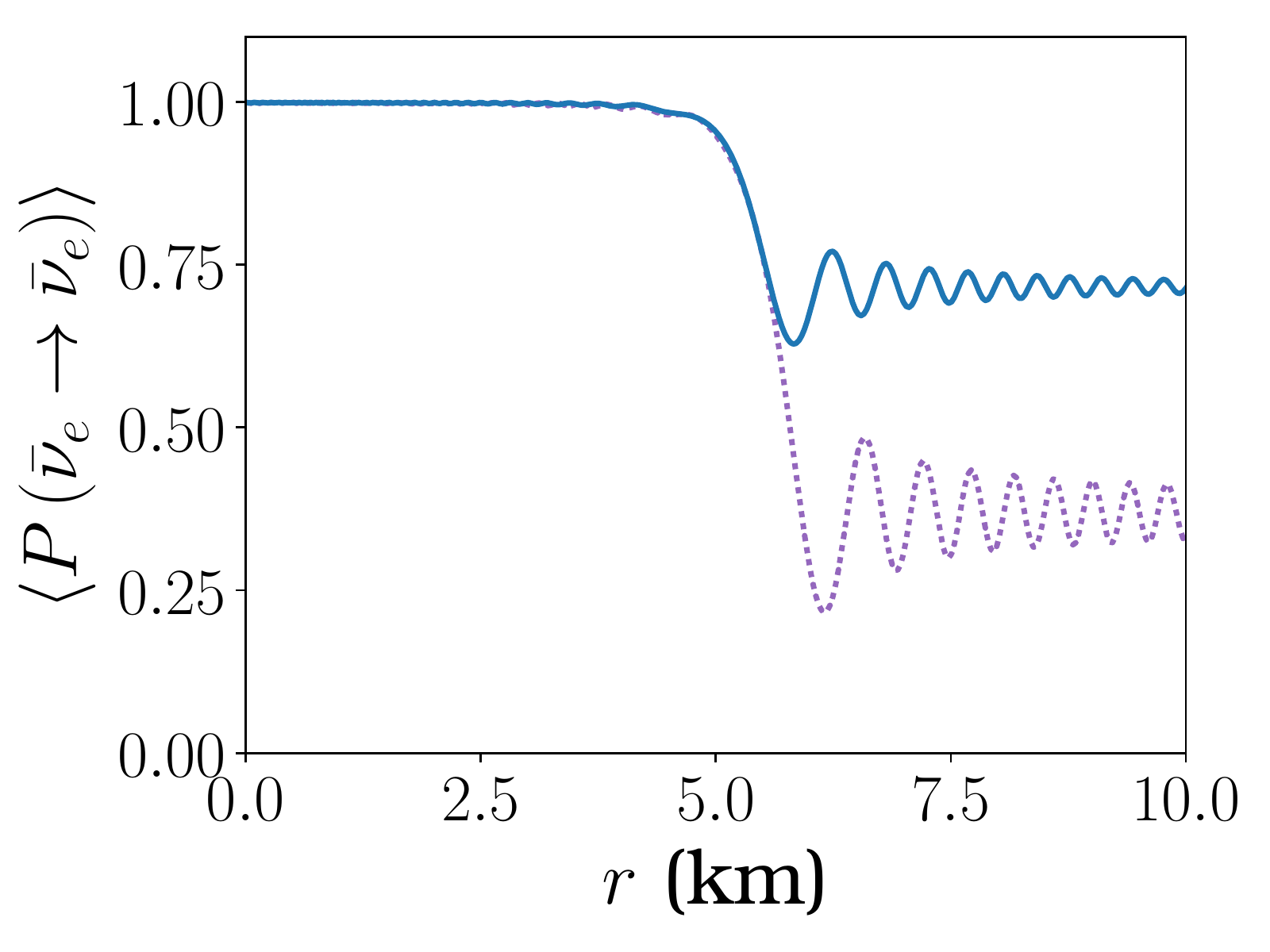}
\caption{Left panel : Difference of the diagonal elements of the total neutrino Hamiltonian, as a function of distance from the emission point, for normal (solid lines) and inverted (dotted lines) mass ordering. The initial parameters are $x_0 =-10$ km, $z_0 =30$ km, and $\theta_q =25^\circ$. 
Middle and right panels :
Averaged survival probabilities  for neutrinos (middle), antineutrinos (right).  The NSI parameters are set to $\delta \epsilon^n = -0.90$ and $\epsilon_0 = 1\times 10^{-4}$.}
\label{fig:modE_90_e-4}
\end{figure*}

\paragraph{Effect of the neutrino mass ordering}
The sign of $\omega_\text{sync}$ changes when going from normal to inverted mass ordering. However, due to the fact that the resonance location almost coincides with the location at which $V_M$ changes its sign, the mass ordering will have little impact on it. In our calculations we have found modifications of the resonance location smaller than $1$ km between normal and inverted mass ordering. 
As for the adiabaticity parameter \eqn{eq:adiabI},  it also depends on $\omega_\text{sync}$ and its derivative. Figure \ref{fig:modE_90_e-4} shows the effect of neutrino mass ordering on the adiabaticity of flavor evolution for a case with $\delta \epsilon^n = -0.90$ and $\epsilon_0 = 1\times 10^{-4}$ where the I resonance is located very close the neutrinosphere, at 5 km.

\subsection{NSI, the MNR and the I resonance}
\begin{figure*}
\includegraphics[width=.3\textwidth]{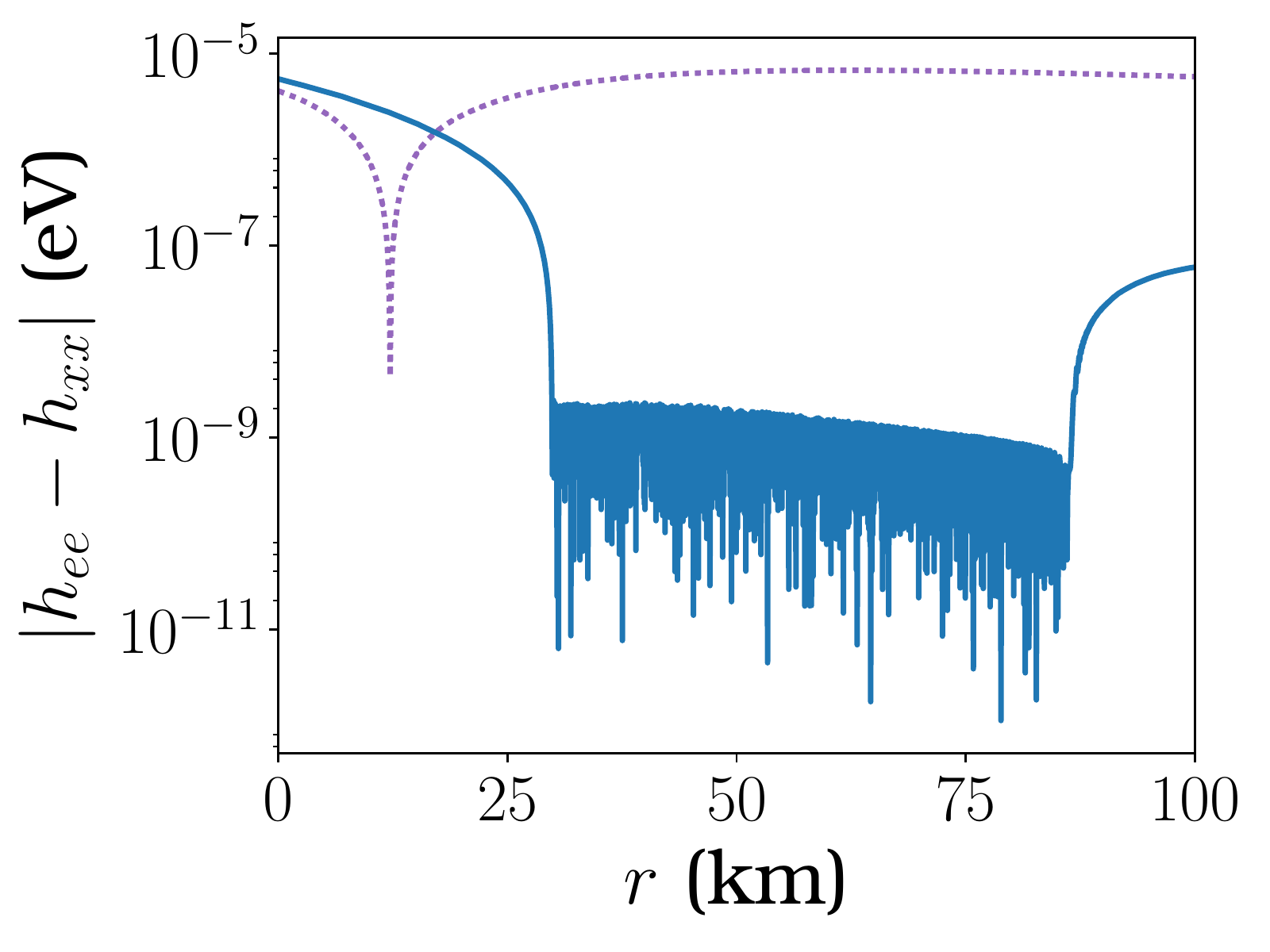}
\includegraphics[width=.3\textwidth]{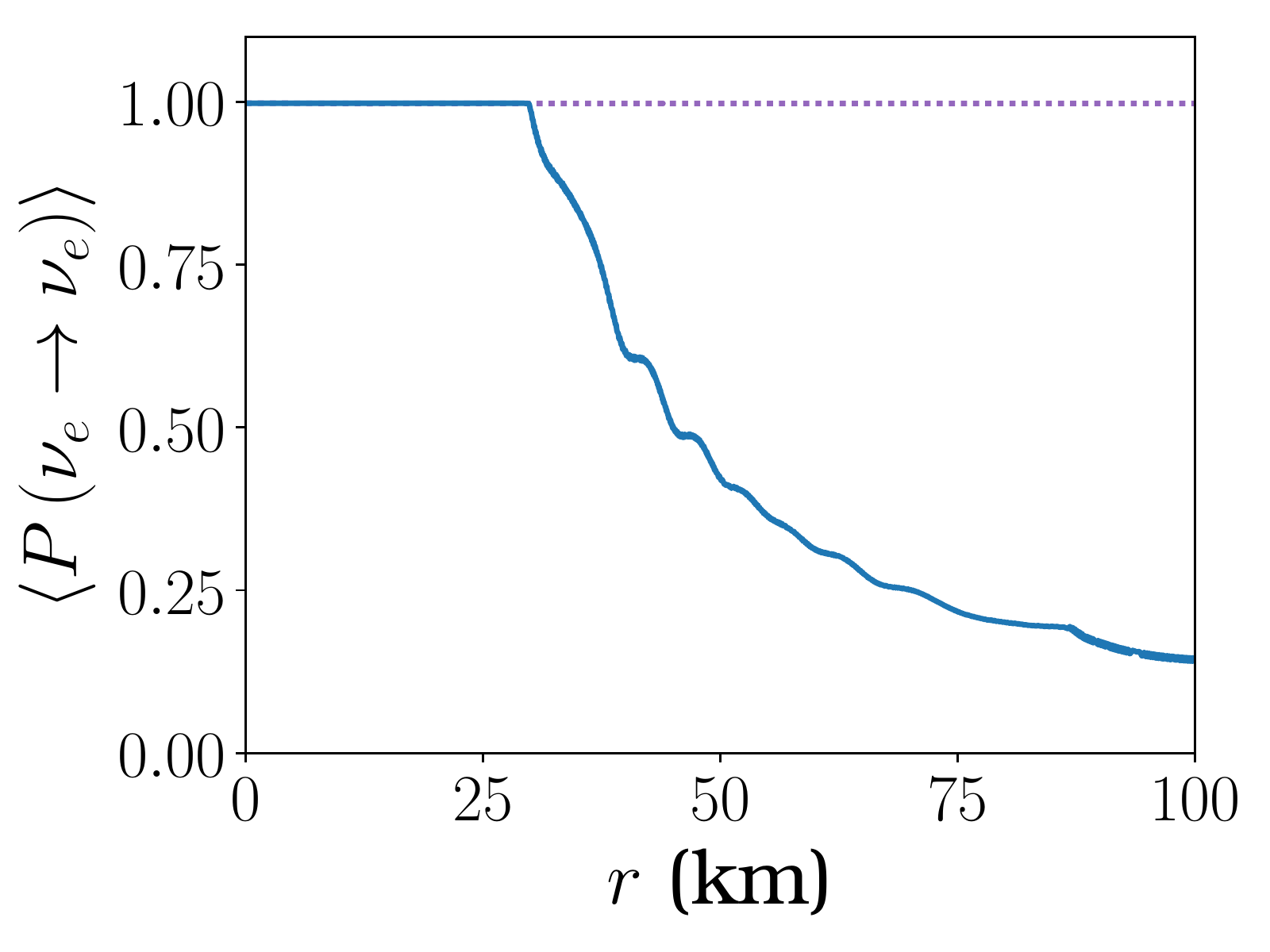}
\includegraphics[width=.3\textwidth]{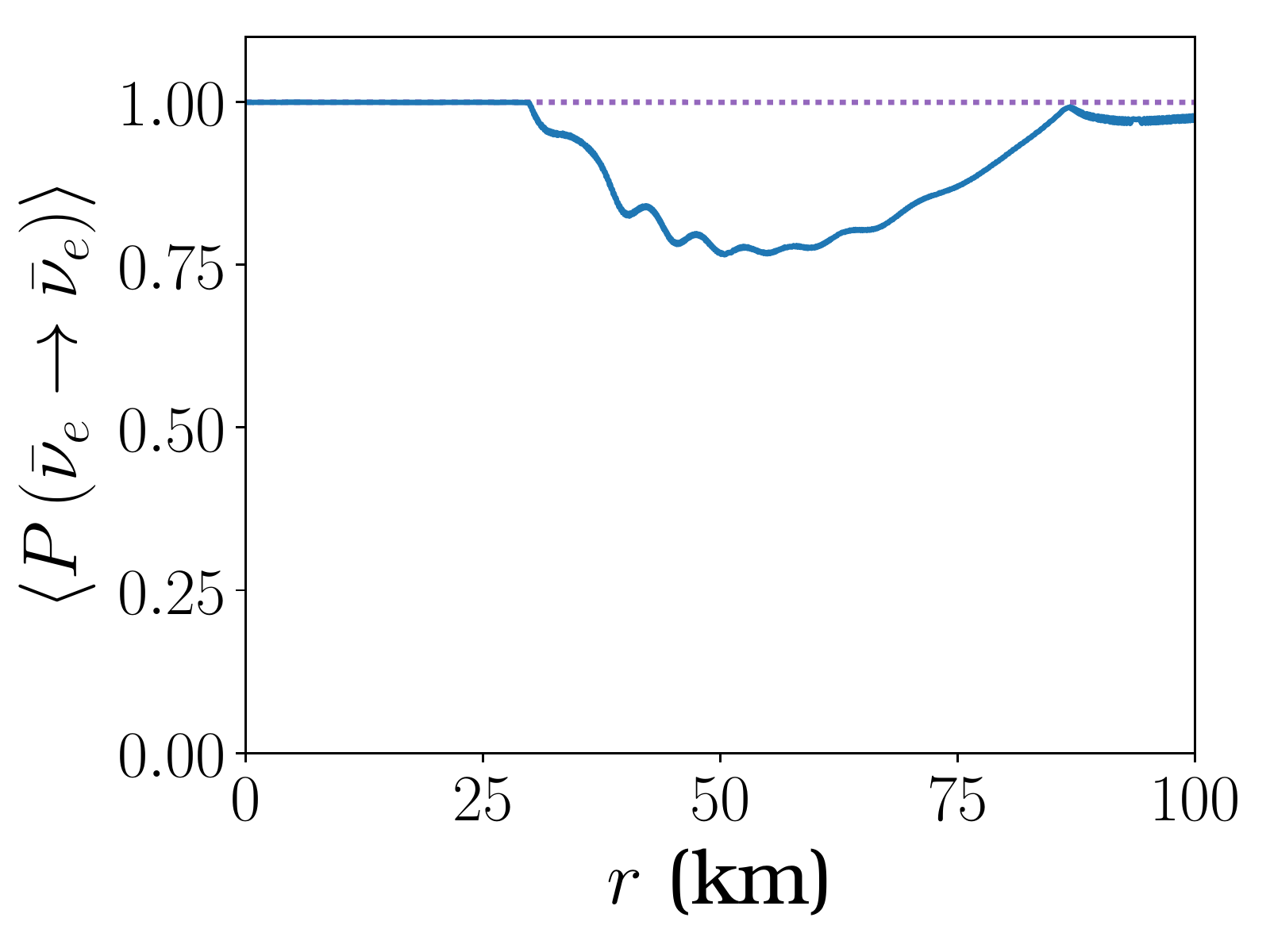}
\caption{Left panel : Difference of the diagonal elements of the total neutrino Hamiltonian, as a function of distance from the emission point, without NSI (dotted line) and with NSI parameters $\delta \epsilon^n = -0.70$ and $\epsilon_0 = 1\times 10^{-4}$ (dotted line). The initial parameters are $x_0 =12$ km, $z_0 =27$ km, and $\theta_q =40^\circ$. 
Middle and right panels :
Averaged survival probabilities  for neutrinos (middle), antineutrinos (right).}
\label{fig:modB_70_e-4}
\end{figure*}

The occurrence of the MNR in BNS might impact $r$-process nucleosynthesis in neutrino-driven winds, as discussed in Ref.\cite{Malkus:2015mda}. 
Two kinds of MNR have been pointed out, either a symmetric one  in which both neutrinos and antineutrinos convert \cite{Malkus:2012ts}, or a standard one where only neutrinos undergo flavor conversion \cite{Malkus:2015mda}. 
The MNR phenomenon is due to a cancellation between the standard matter term  Eq.(\ref{e:hmat2f}) and the neutrino self-interaction Eq.(\ref{e:hexpsa}). This occurs because of the excess of the antineutrino over the neutrino near the disk in the BNS context, compared to the supernova case, that gives a negative sign to the neutrino self-interaction potential $\mu$ (\ref{e:muunosc}). However, Ref.\cite{Stapleford:2016jgz} has shown the presence of NSI can trigger the MNR also in the supernova context. In our numerical investigations, we have observed various NSI effects on the flavor behaviours in presence of MNR. First the existence of NSI can modify the location of the MNR.
Figure \ref{fig:modB_70_e-4} shows that the cancellation between the matter and the neutrino self-interaction terms shifts from 10 km to 30 km when NSI are included. Moreover neutrino evolution turns from completely non-adiabatic to adiabatic, as the the survival probabilities show. By looking at the difference of the neutrino Hamiltonian diagonal elements, one can see that they keep being very small from 30 km to 80 km due to the non-linear feedback that matches the nonlinear neutrino self-interaction contribution to the matter potential \cite{Chatelain:2016xva}.  

\begin{figure*}
\includegraphics[width=.3\textwidth]{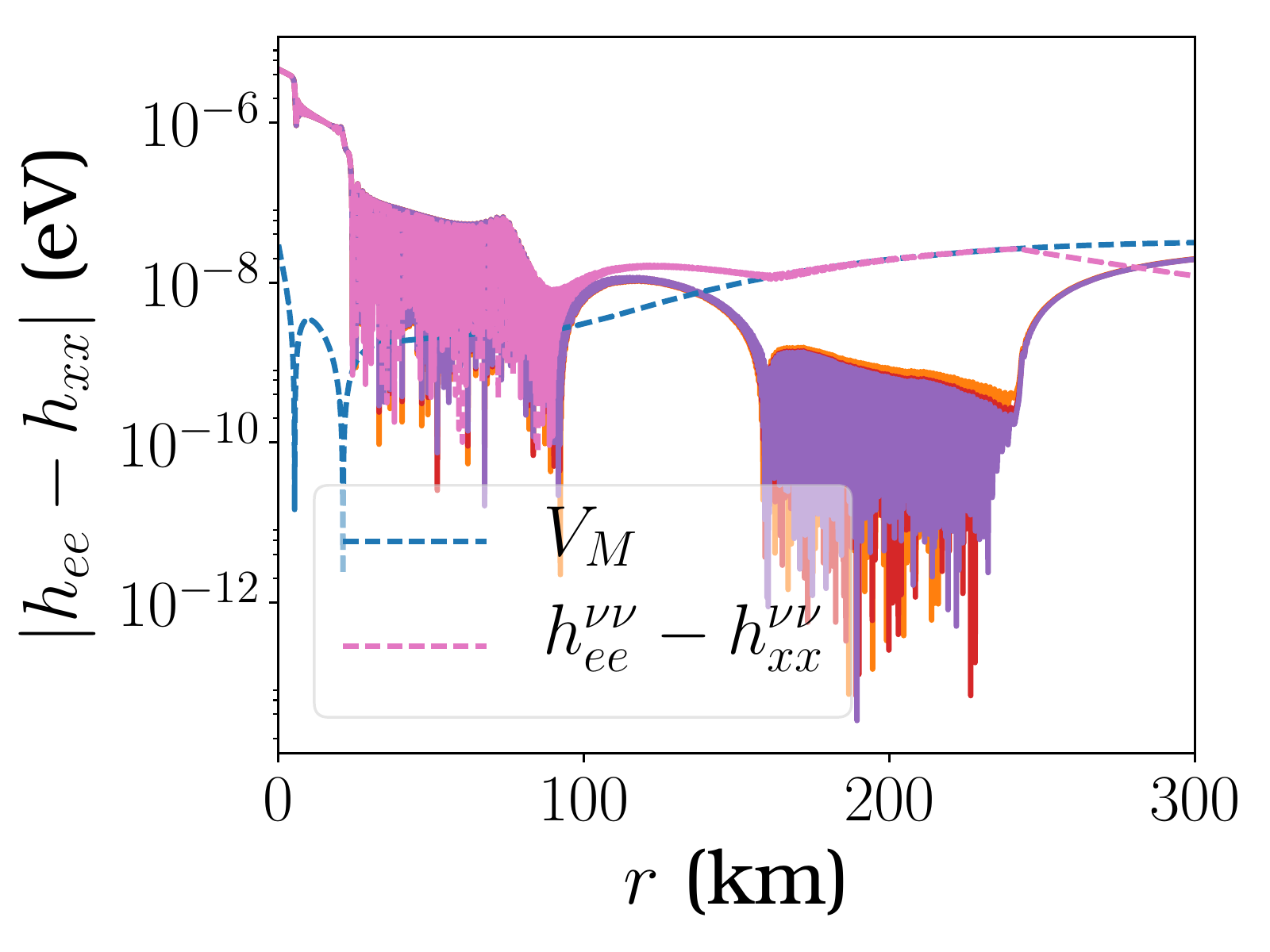}
\includegraphics[width=.3\textwidth]{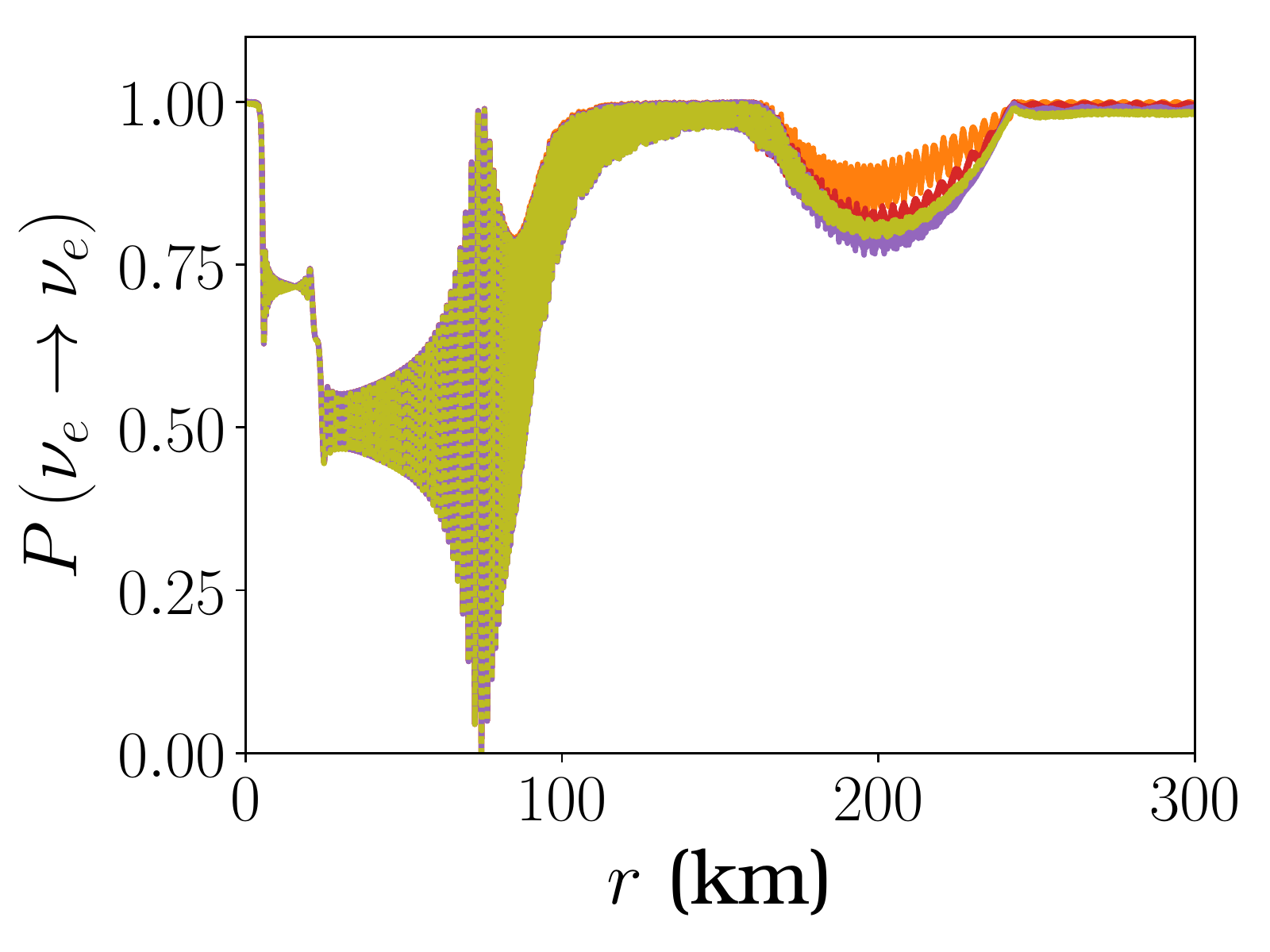}
\includegraphics[width=.3\textwidth]{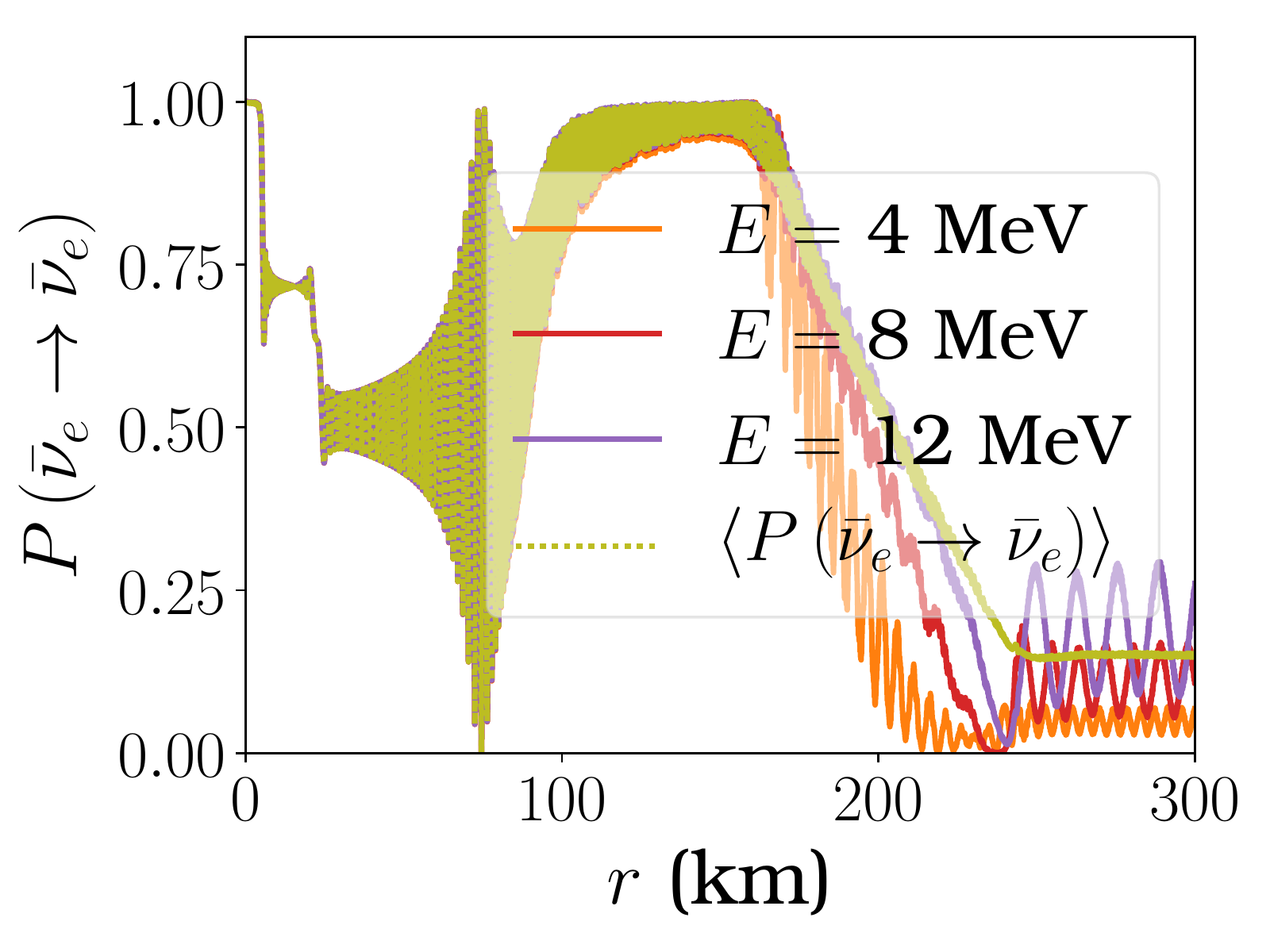}
\caption{Left panel : Difference of the diagonal elements of the total neutrino Hamiltonian (solid line), matter  potential  $V_M$ (dashed line) Eq.(\ref{e:Ires}) in presence of NSI contributions with $\delta \epsilon^n = -0.90$ and $\epsilon_0 = 1\times 10^{-4}$ and self-interaction oscillated potential (dotted line), as a function of distance from the emission point. The initial parameters are $x_0 =-10$ km, $z_0 =30$ km, and $\theta_q =25^\circ$. Middle and right panels : Survival probabilities  for neutrinos (middle), antineutrinos (right). Different energies correspond to different colors, and the averaged probabilities (dotted line) are shows. The slight dependence on the energy is due to the fact that as the MNR occurs further away from the emission point, the difference between the diagonal elements becomes comparable to the vacuum term, which then plays a role. }
\label{fig:modE_90_e-4_full}
\end{figure*}

\begin{figure*}
\includegraphics[width=.3\textwidth]{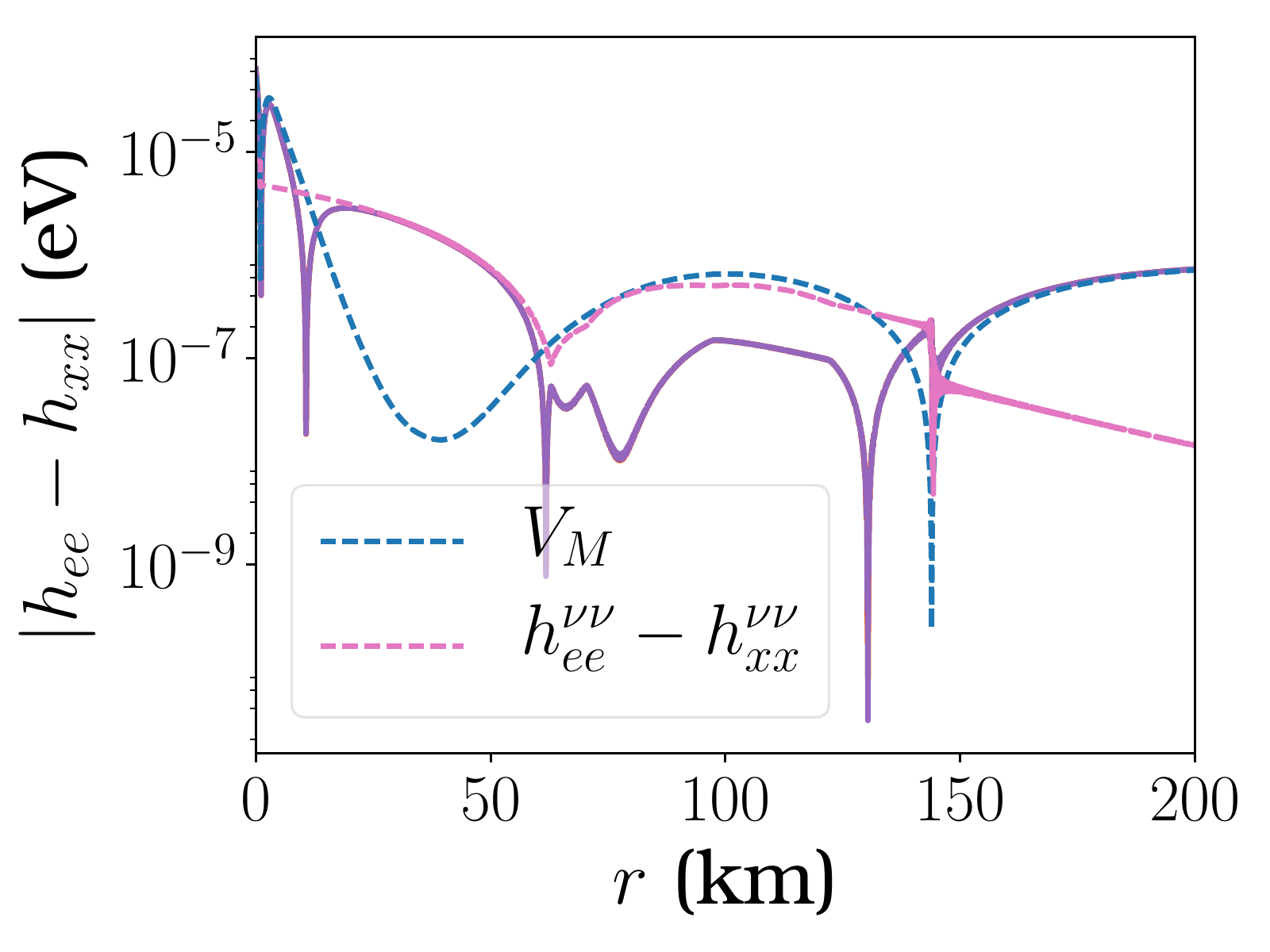}
\includegraphics[width=.3\textwidth]{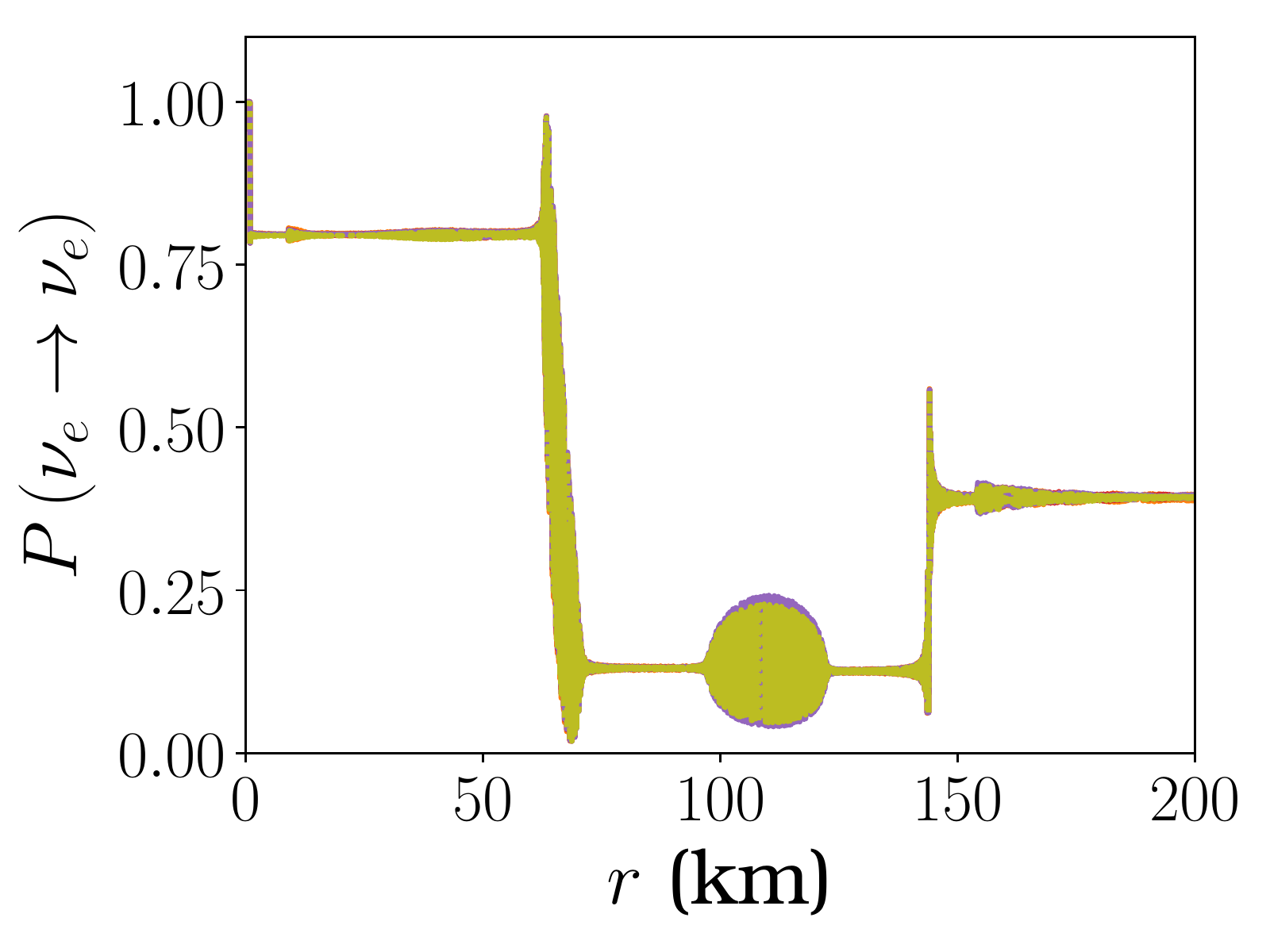}
\includegraphics[width=.3\textwidth]{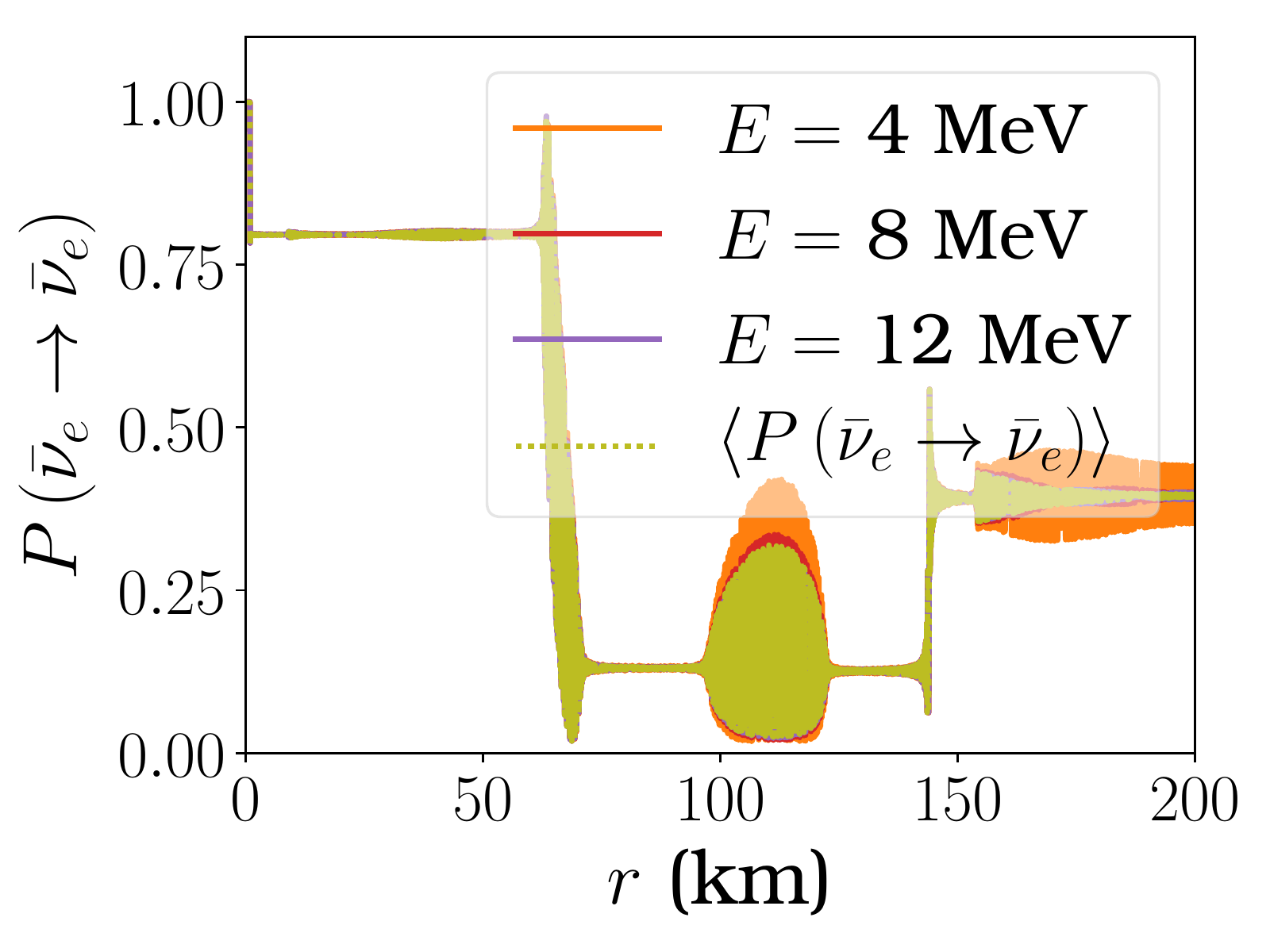}
\caption{Left panel : Difference of the diagonal elements of the total neutrino Hamiltonian (solid line), matter  potential  $V_M$ (dashed line) Eq.(\ref{e:Ires}) in presence of NSI contributions with $\delta \epsilon^n = -0.70$ and $\epsilon_0 = 1\times 10^{-5}$ and self-interaction oscillated potential (dotted line), as a function of distance from the emission point. The initial parameters are $x_0 =-30$ km, $z_0 =20$ km, and $\theta_q =55^\circ$. Middle and right panels : Survival probabilities  for neutrinos (middle), antineutrinos (right). Different energies corresponding to different colors as well as averaged probability (dotted line) are indistinguishable. }
\label{fig:modC_70_e-5}
\end{figure*}

\begin{figure*}
\includegraphics[width=.3\textwidth]{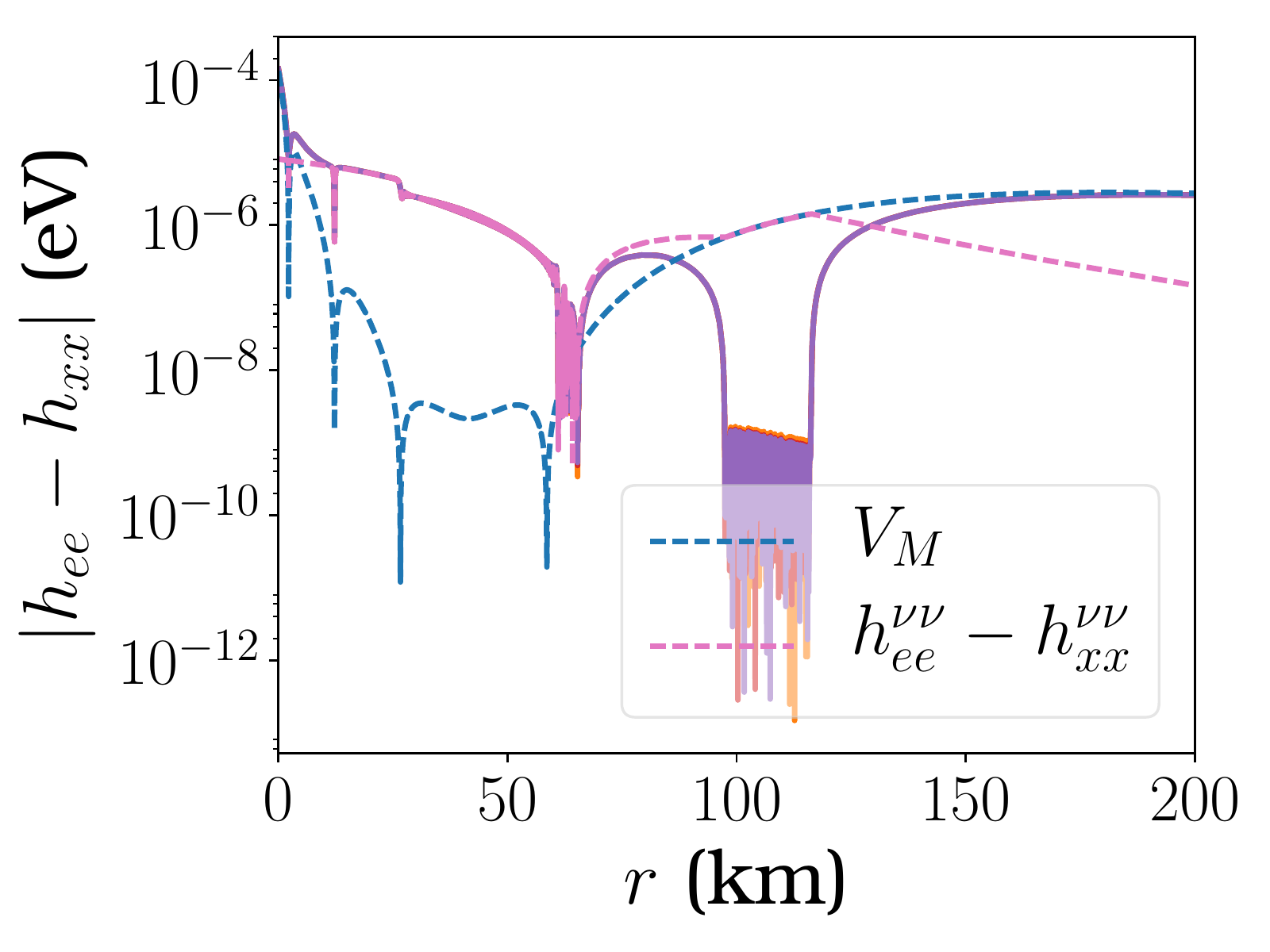}
\includegraphics[width=.3\textwidth]{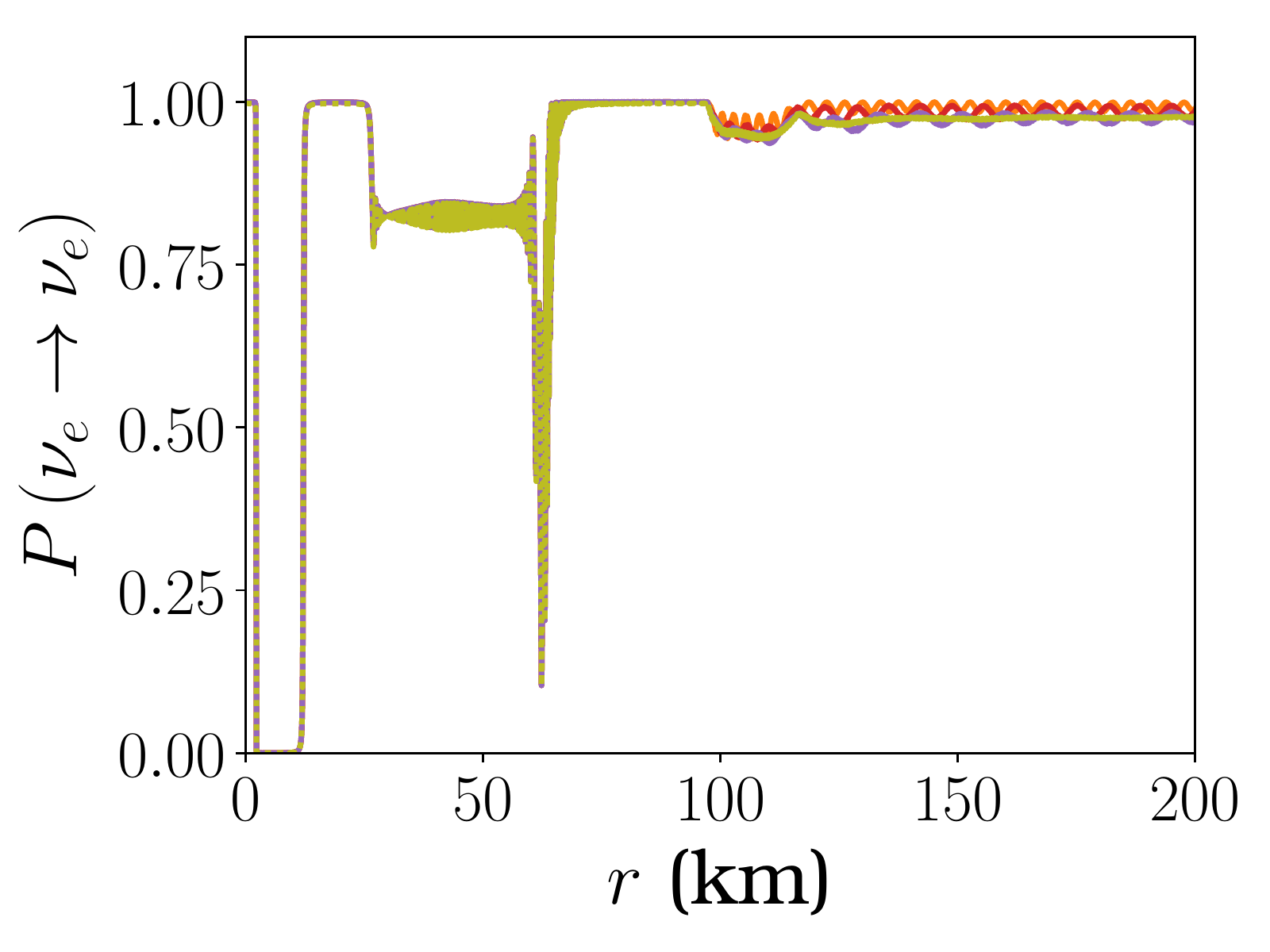}
\includegraphics[width=.3\textwidth]{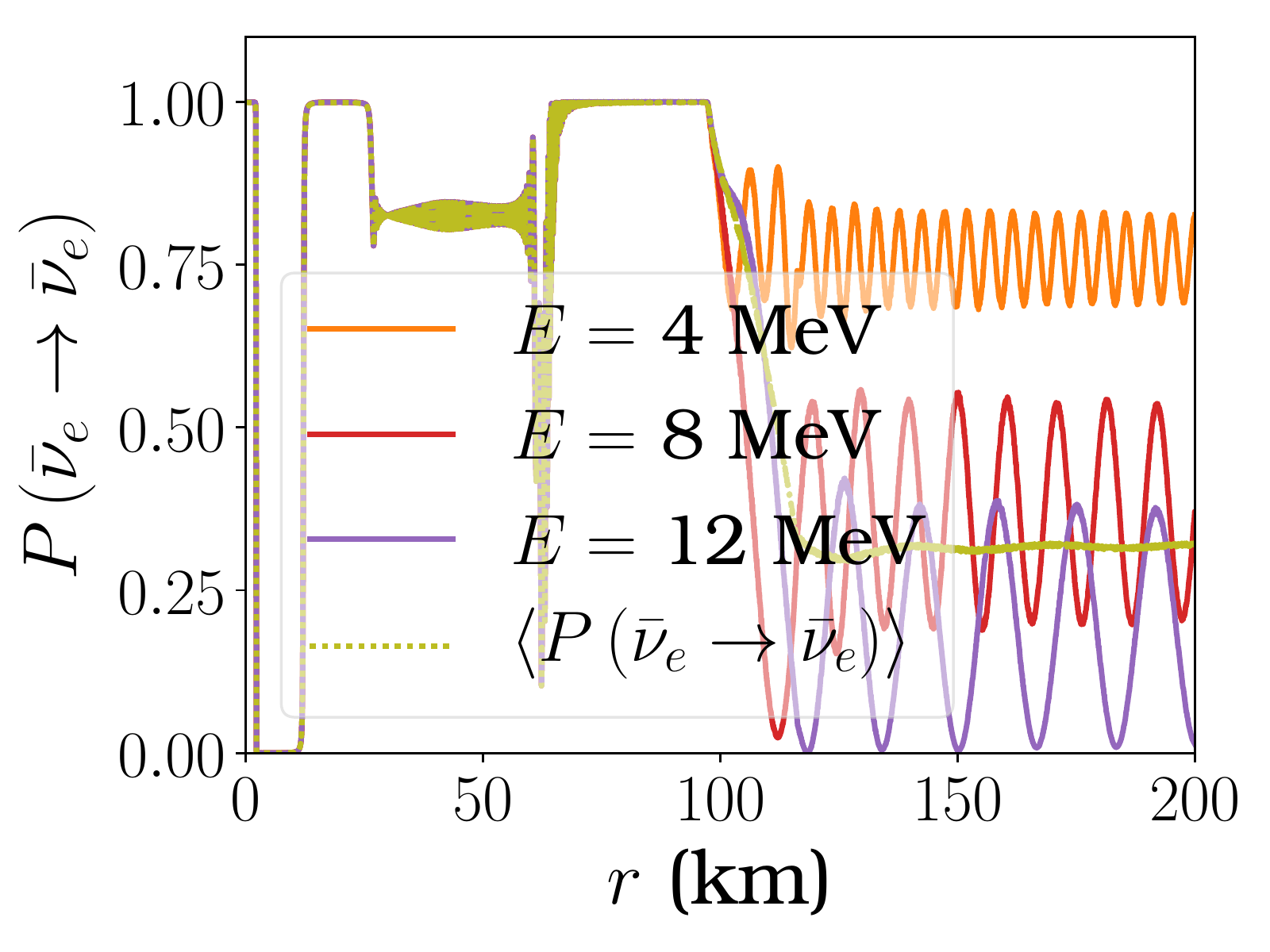}
\caption{Left panel : Difference of the diagonal elements of the total neutrino Hamiltonian (solid line), matter  potential  $V_M$ (dashed line) Eq.(\ref{e:Ires}) in presence of NSI contributions with $\delta \epsilon^n = -0.90$ and $\epsilon_0 = 1\times 10^{-4}$ and self-interaction oscillated potential (dotted line), as a function of distance from the emission point. The initial parameters are $x_0 =-30$ km, $z_0 =20$ km, and $\theta_q =55^\circ$. Middle and right panels : Survival probabilities  for neutrinos (middle), antineutrinos (right). Different energies correspond to different colors, and the averaged probabilities (dotted line) are shown. The slight dependence on the energy is due to the fact that as the MNR occurs further away from the emission point, the difference between the diagonal elements becomes comparable to the vacuum term, which then plays a role.  }
\label{fig:modC_90_e-4}
\end{figure*}

Along numerous trajectories and sets of NSI parameters we have observed an intriguing interplay between the I  resonance, synchronized or not, and the MNR.
Figures \ref{fig:modE_90_e-4_full}, \ref{fig:modC_70_e-5} and \ref{fig:modC_90_e-4}  furnish three examples of such behaviours. 
Figure \ref{fig:modE_90_e-4_full} shows a combination of I resonance and MNR. There are two I resonances, the first at $5$ km which is partially adiabatic, and the second at $21$ km, which triggers a MNR between $20$ km and $100$ km, followed by a second one between $160$ km and $240$ km where the $\bar{\nu}_e$ are converted while $\nu_e$ are not \footnote{Note that this corresponds to  the same parameters as the ones of Figure \ref{fig:modE_90_e-4} with a larger range shown.}. Note that this is in opposition to what the MNR typically creates in the absence of NSI : indeed, without NSI, the MNR tends to lead to flavor conversions for neutrinos while for antineutrinos the evolution is generally non- or partially adiabatic.
In Figure \ref{fig:modC_70_e-5}, an I resonance is located at $2$ km, followed by a non-adiabatic MNR at $12$ km. Then, between $60$ km and $70$ km MNR conversions takes place. Between $100$ km and $125$ km the difference of the diagonal elements stays very small, creating small conversions. Finally, at $144$ km, an I resonance occurs.
The third example of a combination of MNR and I resonances is given in Figure \ref{fig:modC_90_e-4}. This case in point is interesting as it shows four I resonances : the first, located around $2$ km, being a standard one, completely adiabatic, and the other three being synchronized resonances. At $12$ km, the second resonance is also very adiabatic, then the third, at $26$ km creates only partial conversions.  A fourth resonance occurs at $58$ km and produces a short MNR-like cancellation between $60$ km and $66$ km, followed by a MNR between $96$ km and $126$ km. Notice again the peculiar behavior of this MNR, which creates conversions for antineutrinos while the evolution for neutrino is nonadiabatic. \\

\section{Discussion and conclusions}
\begin{figure}[!]
\includegraphics[width=.5\textwidth]{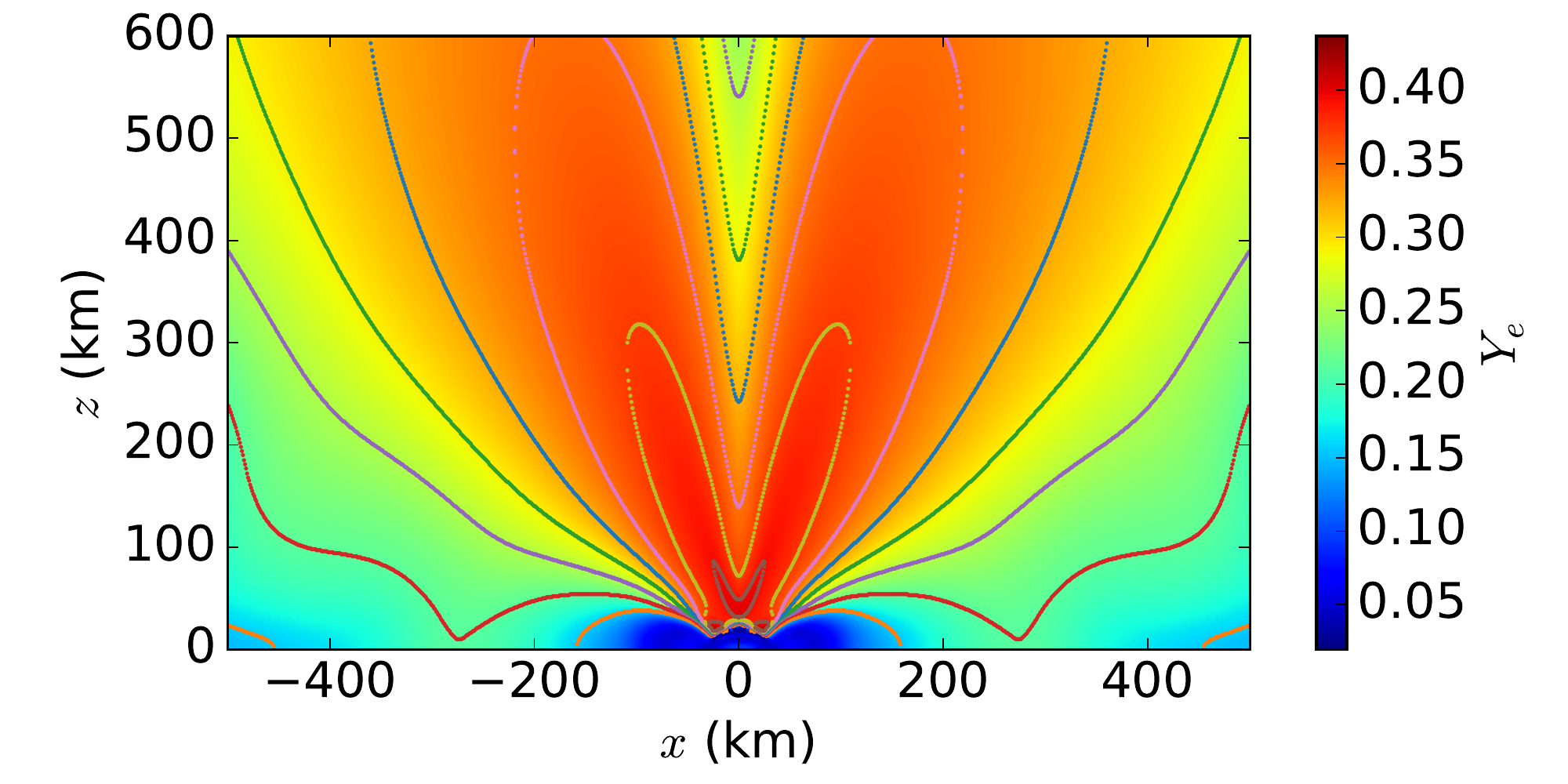}
\caption{Locations where the I resonance condition Eq.\ref{e:Ires} is fulfilled, depending on the NSI parameters $\delta \epsilon^n$. The curves from outside (orange) to inside (brown) correspond to values from  $\delta \epsilon^n = -0.2 $ to $- 0.9 $ in steps of $-0.1$. The $Y_e$ distribution is taken from the BNS simulations of Ref. \cite{Perego:2014fma}.}
\label{fig:Ye}
\end{figure}

In order to assess the role of flavor evolution on nucleosynthesis in neutrino driven winds a self-consistent calculation of the electron fraction modification
coupled to the flavor evolution should be performed, as e.g. the one performed in Ref. \cite{Tamborra:2011is} in core-collapse supernovae. First steps in this direction are presented in Refs.\cite{Malkus:2012ts,Malkus:2015mda}\footnote{Note that in these calculations are not fully consistent since the feedback effect of the modified electron fraction on the probabilities is not included}. However the trajectory dependence on the abundances and investigations without the {\it ansatz} (\ref{e:st}) need to be performed. Such studies go beyond the scope of the present work. Figure \ref{fig:Ye} shows the I resonance location according to Eq.(\ref{e:Ires}) in the dimensional space. One can see that such resonance can occur close to the neutrinosphere and for a large set of NSI parameters. Obviously, for the cases where only the matter term matters, the resonance location would keep unchanged if the {\it ansatz} (\ref{e:st}) is relaxed. Using the at-equilibrium $Y_e$ as a reference, one would expect that the $Y_e$ value should be increased by the presence of I resonances since the $\nu_e$ and $\bar{\nu}_e$ conversion to $\nu_x$ and $\bar{\nu}_x$ respectively brings the former to have the average energies of the latter. However, the at-equilibrium $Y_e$ is certainly not a good reference for the conditions encountered very close to the neutrinosphere.
Only a consistent calculation of $Y_e$ modification including the feedback on the probabilities and the full angular dependence of the neutrino emission would tell us how much flavor evolution impacts the electron fraction.

In the present work,  we have investigated the role of nonstandard matter-neutrino interactions within 2$\nu$ flavor framework. In particular we have included the electron-tau couplings for which 
current bounds from scattering and oscillation experiments are still rather loose. By solving the mean-field Liouville Von-Neumann equations along a large ensemble of trajectories, we have uncovered aspects of NSI impact on flavor evolution and in particular on the I resonance and the MNR.
First, we have shown the conditions for the I resonance are met in this kind of setting, based on detailed BNS simulations, when the matter term dominates over the self-interaction contribution to the neutrino Hamiltonian. Then, we have uncovered the role of the neutrino self-interaction term and shown that the I resonance can be a synchronized MSW effect the self-interaction potential dominates over the matter one. The synchronized precession frequency, depending on by the self-interaction potential, matches the resonance condition when the total matter term becomes very small. This mechanism has been dismissed in previous investigations. Note that in Ref.\cite{Frensel:2016fge} a synchronized MSW effect is observed when, on the contrary, the self-interactions become very small. Second, for the MNR we have shown that NSI modify little the resonance location while the adiabaticity can 
be significantly changed. Third we have shown complex situations where MNR, I and synchronized I combine producing
intriguing flavor patterns. 
 
The contribution of heavy elements nucleosynthesis in BNS  is an open question.
The discovery of gravitational waves \cite{Abbott:2016blz}, 
the determination of the BNS rate from the LIGO and Virgo collaborations and the kilonova observation\cite{TheLIGOScientific:2017qsa}  bring crucial information
to the longstanding puzzle of the origin of $r$-process nuclei. To answer this question,
 one needs to assess the BNS rate as well as the amount of elements from each individual event. 
The kilonova observation constitutes an experimental proof that 
heavy elements are indeed produced in BNS. In such sites 
nucleosynthetic abundances can be produced in the dynamical ejecta and neutrino-driven winds (see e.g. \cite{Martin:2015hxa, Just:2014fka}). Which elements are
produced in each needs to be assessed.
In this respect it is necessary to determine if and under which conditions flavor evolution 
takes place as well as its influence on nucleosynthetic abundances. The present work provides insight to progress in this direction, also in presence of new physics as nonstandard interactions that might be discovered in the future.

\begin{acknowledgments}
\noindent
We thank Maik Uwe Frensel and Albino Perego for useful discussions. The authors acknowledge support
from "Gravitation et physique fondamentale" (GPHYS) of the {\it Observatoire de Paris}.
\end{acknowledgments}

\end{document}